\documentclass[aps,prb,groupedaddress,superscriptaddress,twocolumn]{revtex4-2}
\usepackage{amsmath,graphicx}
\usepackage{float}
\usepackage{subfigure}
\usepackage{color}
\usepackage{braket}
\usepackage{mathrsfs}
\usepackage{appendix}
\usepackage{amssymb}

\begin{document}
\title{Wavefront dislocations in graphene systems revealed by transport measurement}
\author{Yu-Chen Zhuang}
\affiliation{International Center for Quantum Materials, School of Physics, Peking University, Beijing 100871, China}
\affiliation{CAS Center for Excellence in Topological Quantum Computation, University of Chinese Academy of Sciences, Beijing 100190, China}

\author{Qing-Feng Sun}
\email[]{sunqf@pku.edu.cn}
\affiliation{International Center for Quantum Materials, School of Physics, Peking University, Beijing 100871, China}
\affiliation{CAS Center for Excellence in Topological Quantum Computation, University of Chinese Academy of Sciences, Beijing 100190, China}
\affiliation{Hefei National Laboratory, Hefei 230088, China}

\date{\today}

\begin{abstract}
The wavefront dislocation is an important and ubiquitous phenomenon in wave fields. It is closely related to the phase singularity in a wave function. Some recent studies have verified that the wavefront dislocations in the local density of states (LDOS) map can well manifest the intrinsic topological characteristics in graphene and some topological systems. Different from these previous schemes, we raise a transport method to measure such wavefront dislocations in monolayer and Bernal-stacked bilayer graphene. Combining analytical analysis and numerical calculation, we find phase singularities naturally appear in the transmission coefficients between different sublattices, due to the intervalley interference on the electron propagating paths. These phase singularities could contribute wavefront dislocations in the conductance map. Additionally, in bilayer graphene, the wavefront dislocations are found to remain robust even though the tip is coupled to multiple sublattices. Biased bilayer graphene is also explored. Our scheme provides a new transport routine to explore valley-related topological properties of materials.
\end{abstract}

\maketitle	
\section{\label{sec1} Introduction}
In a wavefield, some additional wavefronts (i.e. surface of constant phase) may appear in a certain area associated with a topological defect, which are known as wavefront dislocations. Following the pioneering work of Nye and Berry in $1974$ \cite{Nye}, this fundamental and ubiquitous phenomenon
can in principle emerge in any wavefield irrespectively of its physical nature or its dispersion relation. In mathematics, it originates from a phase indetermination in which the amplitude of wave vanishes.
One example is about Aharonov-Bohm effect \cite{Berry, Berry2}, where the electron wave could possess wavefront dislocations on the flux line of the impenetrable cylinder with a magnetic flux. Since the phase singularities are recognized as important features of all waves, the researches on wavefront dislocations have spread over various domains of physics, from astronomy \cite{Mawet}, oceanic tides \cite{Berry3}, sounds \cite{Nye, Melde, Jim}, to optics \cite{Padgett, Rafayelyan} and fluids \cite{Berry, Berry3, Karjanto}. Especially in optics, it gives birth to a new research branch known as singular optics \cite{Soskin,Dennis}.

Due to wave-particle duality, electrons can also be regarded as waves.
In the field of condensed matter physics, it is recently found that wavefront dislocations could closely link the topological defect of waves with the topological properties of materials. This topology is associated with some topological indices, which are featured by the phase singularities of bulk electron wave functions \cite{Berry4, Xiao, Hasan, Delplace, Li}. Thus, through measuring wavefront dislocations originated from the phase singularity, the topological properties of this system can be manifested. For example, in monolayer and bilayer graphene \cite{Dutreix,Zhang3, Zhang}, by introducing a single atom vacancy or a hydrogen atom chemisorbed on a carbon site, dislocations in charge density modulations (i.e. Friedel oscillations \cite{Friedel}) induced by intervalley scatterings can be observed in the scanning tunneling microscopy (STM). The number of wavefront dislocations characterizes the (local) Berry phase signatures. In 1D SSH chain \cite{Dutreix2, Dutreix3}, the wavefront dislocations in the local density of states (LDOS) map of standing wave pattern could present a direct evidence of bulk topological transition.

\begin{figure}[ht]
	\includegraphics[width=0.95\columnwidth]{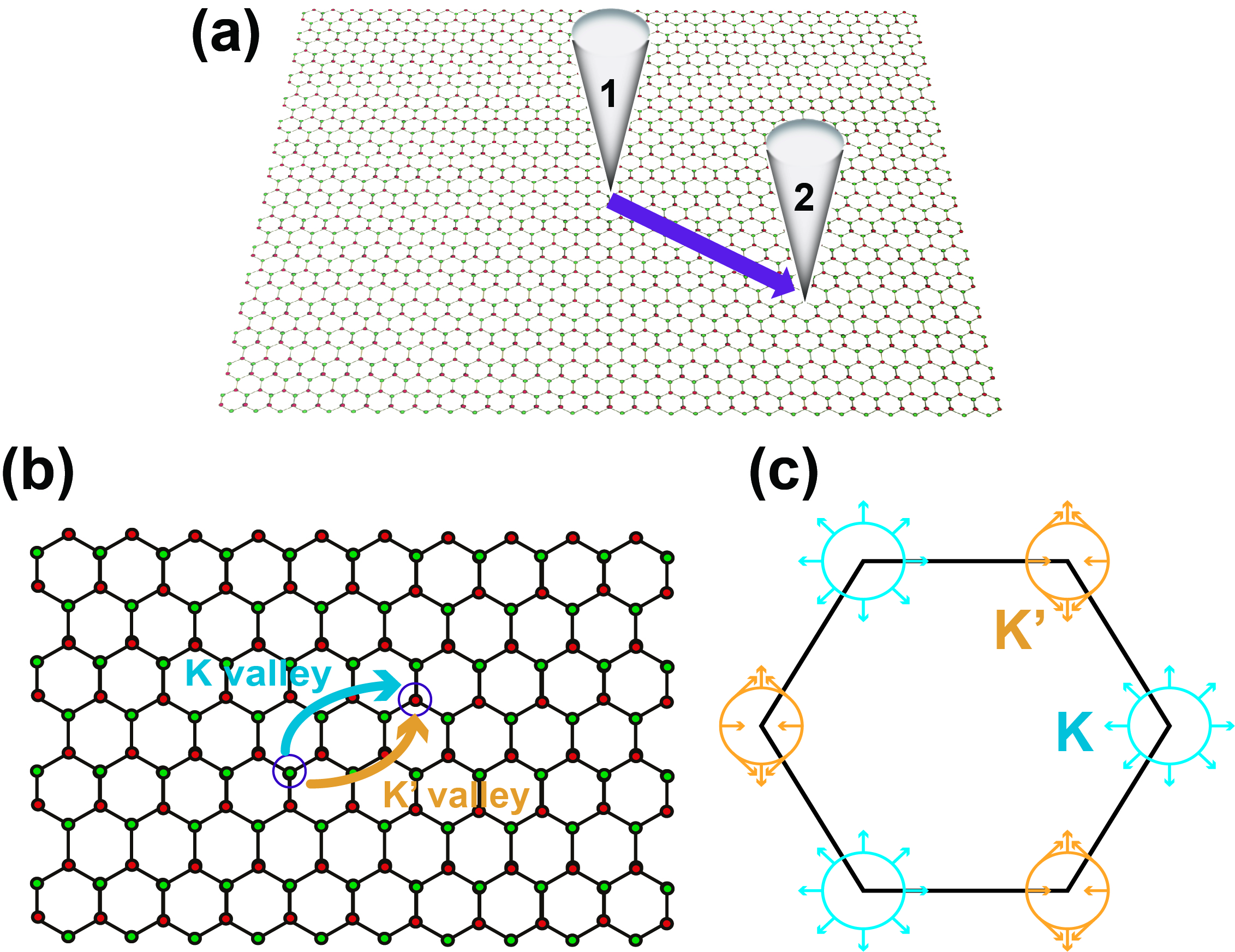}
	\centering 
	\caption{(a) The schematic of the transport measurement via the dual-probe STM setup on the monolayer graphene. (b) The intervalley interference on the electron propagating path between two sublattices. (c) Pseudospin-momentum locking texture along the Fermi surface of monolayer graphene.}
	\label{FIG1}
\end{figure}

The LDOS fluctuations are attributed to
the quantum interference between the incident and scattering wave induced by a defect or boundary in the material \cite{Clark, Chen, Chen2}. If the incident wave function and scattering wave function each carries a different geometric phase, an additional geometrical phase shift could enter the LDOS fluctuations as a topological phase singularity \cite{Dutreix2}. Surrounding this phase singularity along a closed loop, a phase accumulation corresponding to a quantized topological charge will contribute to additional wavefronts \cite{Dutreix}. Except for LDOS, quantum interferences could also influence the transport process where electrons propagate from the source to the drain. So, is there a way to exhibit such wavefront dislocations by the transport measurement?

\begin{figure}
	\includegraphics[width=0.5\textwidth]{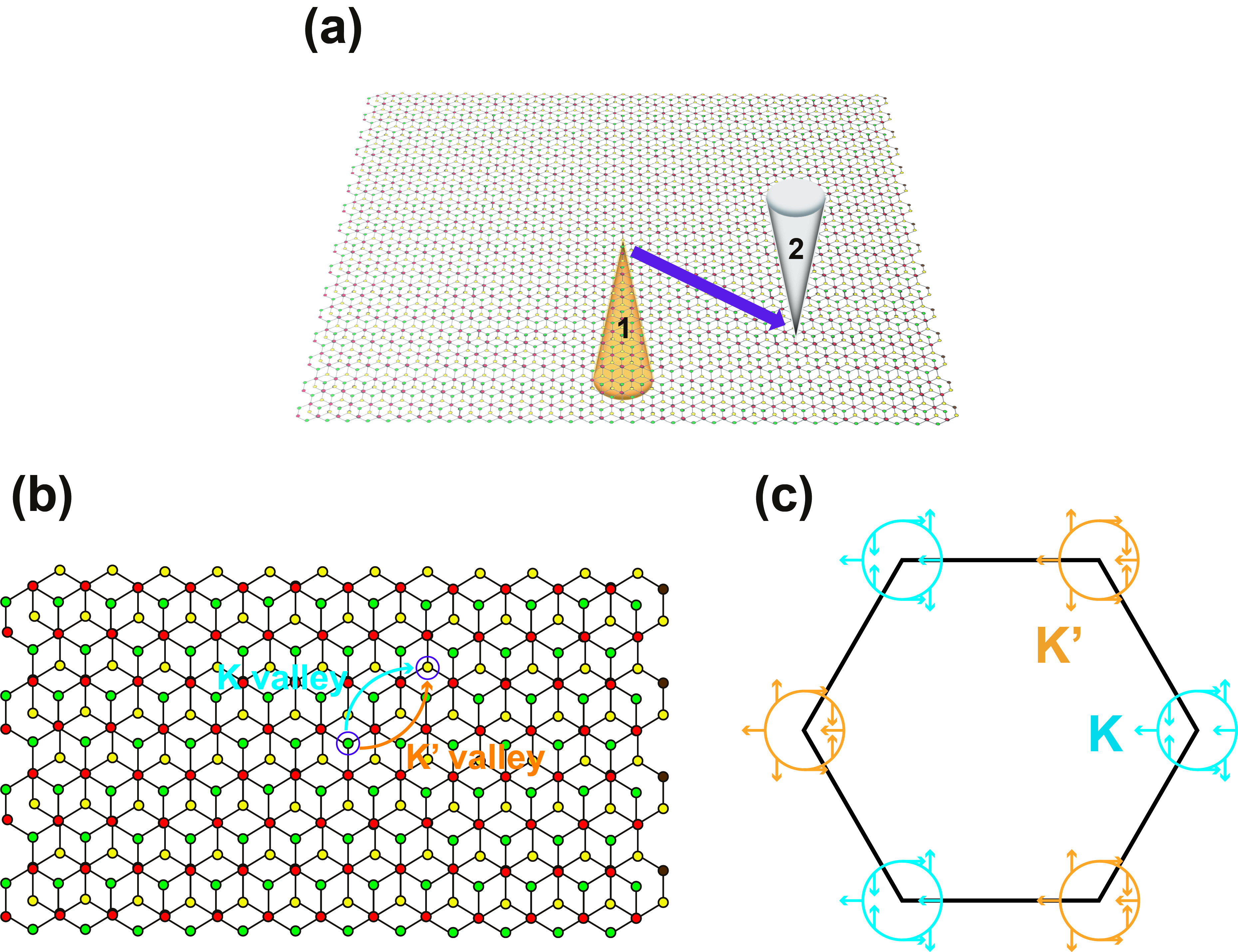}
	\centering 
	\caption{(a) The schematic of the transport measurement via the dual-probe STM setup on the Bernal-stacked bilayer graphene. The fixed probe is placed under the bilayer graphene and connected to the sublattice of the bottom layer. The scanning probe is placed upon the bilayer graphene and connected to the sublattice of the top layer.  (b) The intravalley interference on electron propagating paths between two sublattices. (c) Pseudospin-momentum locking texture along the Fermi surface of Bernal-stacked bilayer graphene.}
	\label{FIG2}
\end{figure}

In this paper, we try to propose one scheme based on the transport measurement to detect wavefront dislocations in graphene systems.
In previous experiments, transport methods usually manifest the topological properties of systems through a quantized two-terminal or Hall conductance \cite{Zhang2, Fang, konig, Zhuang, Law, Mao}. However, in our scheme, a conductance space map is necessary to be obtained. One promising solution is a dual-probe STM experiment with one fixed STM probe and one scanning STM probe \cite{Nakayama, Eder, Settnes, Settnes2}.
Since the graphene has a long mean free path \cite{Mayorov,Rickhaus, Gunlycke}, the interference effects are not washed out by dephasing when two STM tips are placed within a length scale \cite{Rickhaus, Clark}. Both monolayer graphene and (biased) Bernal-stacked bilayer graphene are explored in our scheme. For the monolayer graphene, the fixed STM tip is connected to one sublattice and the scanning probe selectively scans the other sublattices around the fixed STM tip [Fig.~\ref{FIG1}(a)]. For the (biased) Bernal-stacked bilayer graphene, the fixed STM tip is connected to one sublattice of the bottom sheet and the scanning probe selectively scans the other sublattices on the top sheet around the fixed STM tip [Fig.~\ref{FIG2}(a)]. In both cases, due to the intervalley interference on the electron propagating path, wavefront dislocations emerge in the conductance map, reflecting the information of (local) topological indices. In fact, not limited to the dual-probe STM experiment, we emphasize that our scheme could be adapted to more similar systems.

Our paper is organized as follows. In Sec. \ref{sec2}, we use the low-energy Hamiltonian of the monolayer graphene combined with nonequilibrium Green's functions to analyze the transmission coefficients and the conductance between two sublattices. Using the tight-binding model, we numerically obtain the transmission coefficient/conductance maps and find they are well consistent with our analysis. In Sec. \ref{sec3}, the same analysis and calculation procedure is reproduced on the Bernal-stacked bilayer graphene. We demonstrate that the wavefront dislocations are still robust even if the STM tip is coupled to multiple sublattices. This extends the practicality of our theoretical scheme. The case for biased bilayer graphene is also studied. In Sec. \ref{sec4}, we conclude a brief summary and give a discussion about the experiment realization.

\section{\label{sec2} The monolayer graphene}
In this section, we try to explore the wavefront dislocations in the dual-probe STM experiment of the monolayer graphene, as shown in Fig.~\ref{FIG1}(a).
We first assume that each STM probe is only coupled to single $A/B$ sublattice, which is in principle possible in the experiments in view of STM atomic resolution. The case that the STM probes couple to multiple sublattices will be studied in next section.
In fact, our theoretical proposal is not only limited on dual-probe STM experiment, but also applies to other similar systems, like the photonic graphene (see details about experiment implementations in Sec. \ref{sec4}).

\subsection*{A. Analytical analysis}
The massless Dirac fermion in monolayer graphene
can be demonstrated by the low-energy continuum Hamiltonian \cite{Castro,Dutreix}
\begin{equation}
	\begin{split}
		H_{\xi}^{M}(\vec{q}) &= \hbar v_{f}\begin{pmatrix}
			0 & \xi q_{x}-iq_{y}  \\
		    \xi q_{x}+iq_{y} & 0
			\end{pmatrix} \\
			&= \hbar v_{f}\begin{pmatrix}
				0 & \xi q e^{-i \xi \theta_{q}} \\
				\xi q e^{i \xi \theta_{q}} & 0
				\end{pmatrix} \\
	\end{split}
\label{Eq1}
\end{equation}
which is written in the basis of sublattices $\{A,B\}$. Here, the superscript M in the Hamiltonian as well as the following Green's functions denotes the monolayer graphene. $\xi= \pm 1$ denote the valley indices $K$ and $K'$.
$v_{f}$ is the Fermi velocity. $\vec{q}$ is the momentum vector relative to the $K$ or $K'$.
$\theta_{q}=\arctan(q_{y} / q_{x})$ is the polar angle of electrons with the momentum $\vec{q}=(q_{x},q_{y})$.
The eigenvalues of this low-energy Hamiltonian are $E_{\xi}^{\pm}(q) =\pm \hbar v_{f} q$
with the corresponding eigenvectors $\psi_{\xi}^{\pm}(q)=\frac{1}{\sqrt{2}}\binom{1}{\pm \xi e^{i\xi \theta_{q}}}$.
The eigenvectors define pseudospin vectors in the sublattice Bloch space $\langle \psi_{\xi}^{\pm}(q)\vert \vec{\sigma} \vert \psi_{\xi}^{\pm}(q)\rangle =\pm \xi ( \cos \xi \theta_{q}, \sin \xi \theta_{q},0)$, which has the
pseudospin-momentum locking. Here $\vec{\sigma}$ is the vector of Pauli matrices acting on the sublattice space expanded by the basis $\{A,B\}$.
The pseudospin-momentum locking textures are shown in Brillouin zone
for different valleys in Fig.~\ref{FIG1}(c).
Along a closed Fermi surface, the pseudospin will rotate by $2\pi$ yielding a $W=\xi= \pm 1$ winding number and $\gamma=W\pi=\xi\pi$ Berry phase  \cite{Castro,Park,Hou,Liu}.

To analyze the transport properties in the monolayer graphene,
we calculate the bare retarded Green's function matrix with $\mathbf{G}^{0,M}_{\xi}(\omega,\vec{q}) = (\omega+i0^{+}-H_{\xi}^{M}(\vec{q}))^{-1}$ \cite{Dutreix}. It reads ($\omega$ is the tip biased energy)
\begin{equation}
	\begin{split}
		\mathbf{G}_{\xi}^{0,M}(\omega,\vec{q}) = \frac{1}{\omega^{2}-(\hbar v_{f}q)^{2}}\begin{pmatrix}
			\omega &\hbar v_{f} \xi q e^{-i\xi \theta_{q}}  \\
		    \hbar v_{f} \xi q e^{i\xi \theta_{q}} & \omega
			\end{pmatrix}. \\
	\end{split}
\label{Eq2}
\end{equation}
To do Fourier transformation, we use the integral \cite{Dutreix4,Zhang}, 
\begin{equation}
	\iint \frac{d^{2}q}{(2\pi)^{2}}\frac{q^{2m}e^{i\vec{q}\cdot\vec{r}}}{\omega^{2}-q^{2l}}(qe^{i\xi \theta_{q}})^{n} \simeq  -\frac{i^{n+1}\omega^{n/l}e^{i \xi n \theta_{r}}}{4l\omega^{2[1-(m+1)/l]}}H_{n}(\omega^{1/l}r)
	\label{Eq3}
\end{equation}

with the integer number $m$, $n$ and $l$. Note that this equation strictly holds for the case of the monolayer graphene ($l=1$), but is an approximate formula for the case of the multilayer graphene ($l \geq 2$) \cite{Dutreix4}.
The Green's functions of different valleys $G^{0,M}_{K/K'}$ (i.e. $G^{0,M}_{\xi=\pm 1}$ in Eq.~(\ref{Eq2})) in the real space are \cite{Dutreix},
\begin{equation}
	\begin{split}
	\mathbf{G}_{K}^{0,M}(\omega,\vec{r}) &\approx -\frac{\omega e^{i \vec{K} \cdot \vec{r}}}{(2\hbar v_{f})^{2}}\begin{pmatrix}
		iH_{0}(\frac{\omega r}{\hbar v_{f}}) & -H_{1}(\frac{\omega r}{\hbar v_{f}}) e^{-i\theta_{r}} \\ -H_{1}(\frac{\omega r}{\hbar v_{f}}) e^{i\theta_{r}}
		 & iH_{0}(\frac{\omega r}{\hbar v_{f}})
		\end{pmatrix} \\
	\mathbf{G}_{K'}^{0,M}(\omega,\vec{r}) &\approx -\frac{\omega e^{i \vec{K'} \cdot \vec{r}}}{(2\hbar v_{f})^{2}}\begin{pmatrix}
		iH_{0}(\frac{\omega r}{\hbar v_{f}}) & H_{1}(\frac{\omega r}{\hbar v_{f}}) e^{i\theta_{r}} \\ H_{1}(\frac{\omega r}{\hbar v_{f}}) e^{-i\theta_{r}}
		& iH_{0}(\frac{\omega r}{\hbar v_{f}})
		\end{pmatrix}
	\label{Eq4}
    \end{split}
\end{equation}	
where $H_{n}$ denotes the $n$-th order Hankel function of the first kind. The polar angle $\theta_{q}$ related to the sublattice pseudospin mathematically transforms to the polar angle $\theta_{r}=\arctan(\frac{y}{x})$ in the polar coordinate for the vector $\vec{r}=(x,y)$.
It is worth noting that the real space singularity in off-diagonal matrix components is not physical,
since the origin $r=0$ for the bare monolayer graphene is arbitrary. In Friedel oscillations, the origin is determined by a vacancy or an adatom.  In transport measurement, e.g. dual-probe STM measurement,
one of the probes (fixed STM probe) determines the location of the origin, while the other probe (scanning STM probe) scans around in real space.

\begin{figure*}
	\includegraphics[width=2\columnwidth]{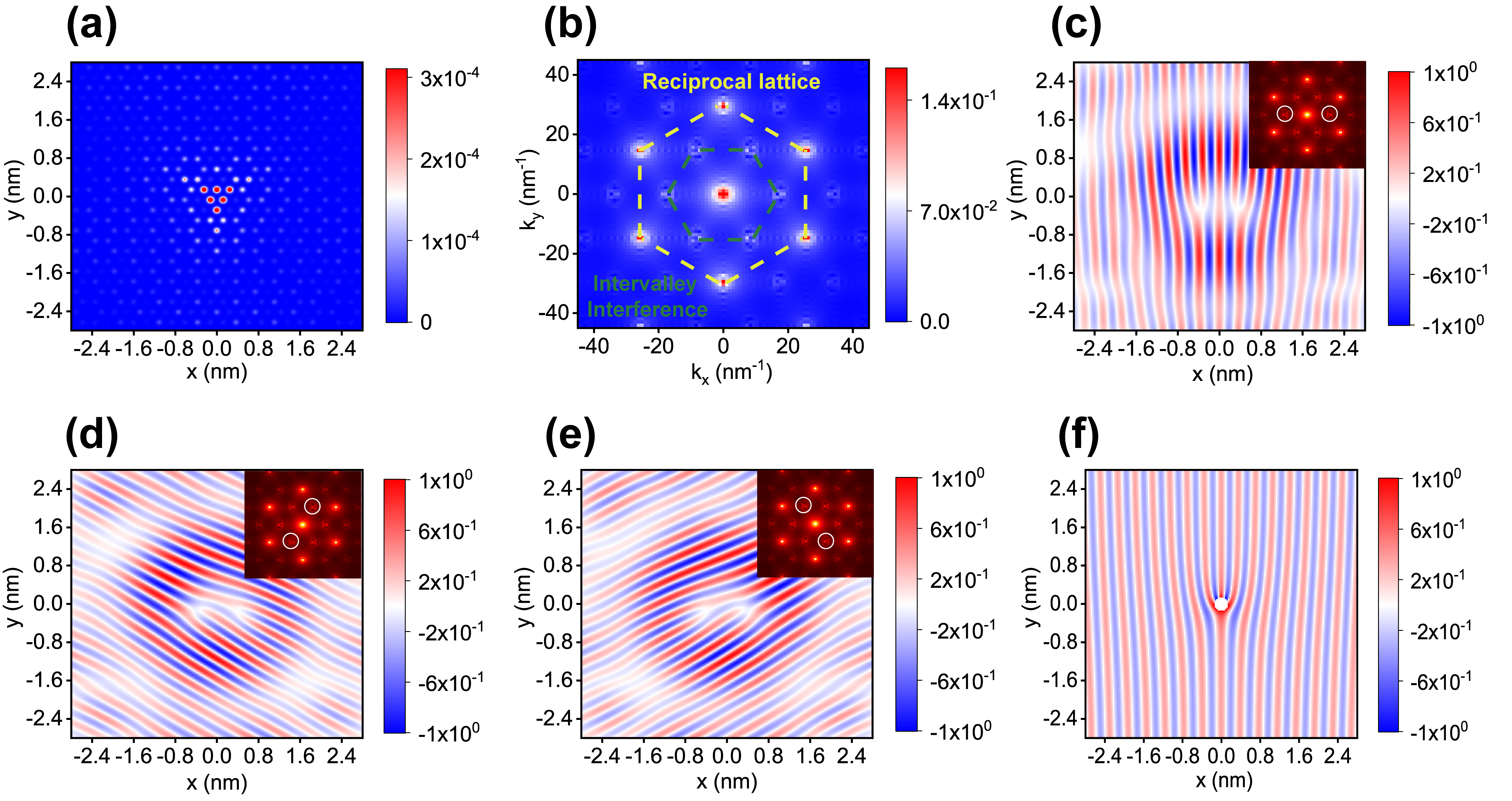}
	\centering 
	\caption{(a) The numerical calculated transmission coefficient map of $T_{2B \leftarrow 1A} (\omega)$. (b) Modulus of the fast Fourier transform (FFT) of the image in (a). The yellow and green dashed lines correspond to reciprocal lattices  and the intervalley interference points respectively. (c-e) The intervalley FFT-filtered images $T_{2B \leftarrow 1A}^{f}(\omega)$ of (b) schematically marked by white circles in the insets for three different directions.
	(f) The plot of the analytical transmission coefficient
	in Eq.~(\ref{Eq10}) which is multiplied by $r$.
	For all the analytical results, the data is not estimated within the range very close to the center $\vec{r}=(x,y)=(0,0)$ due to divergency.}
	\label{FIG3}
\end{figure*}

Our main purpose is to extract the real space singularity resulting wavefront dislocations in the conductance map. In Fig.~\ref{FIG1}(a),
if the fixed probe $1$ is connected to the sublattice $A$ and the scanning probe $2$ scans the sublattice $B$,
the transmission coefficient from probe $1$ to probe $2$ is \cite{Datta, Settnes, Long, Sun}
\begin{equation}
	\begin{split}
	&T_{2B\leftarrow 1A}(\omega,\vec{r}_{1},\vec{r}_{0})\\
	&=Tr[\mathbf{\Gamma}_{2}(\omega,\vec{r}_{1})\mathbf{G}^{M}(\omega,\vec{r}_{1},\vec{r}_{0})\mathbf{\Gamma}_{1}(\omega,\vec{r}_{0})\mathbf{G}^{M,\dagger}(\omega,\vec{r}_{0},\vec{r}_{1})]\\
	&=\Gamma_{B}(\omega,\vec{r}_{1})G_{BA}^{M}(\omega,\vec{r}_{1},\vec{r}_{0})\Gamma_{A}(\omega,\vec{r}_{0})G_{AB}^{M,\dagger}(\omega,\vec{r}_{0},\vec{r}_{1}).
	\label{Eq5}
	\end{split}
\end{equation}
On the second line, the $\mathbf{\Gamma}_{1}(\omega, \vec{r}_{1})$ and $\mathbf{\Gamma}_{2}(\omega, \vec{r}_{2})$ are linewidth function matrices of the probes $1$ and $2$ coupled to graphene lattices. Here $\mathbf{\Gamma}_{1}(\omega, \vec{r}_{1})=\begin{pmatrix}
	\Gamma_{A}(\omega, \vec{r}_{1}) & 0  \\
	0 & 0
	\end{pmatrix}$ and  $\mathbf{\Gamma}_{2}(\omega, \vec{r}_{2})=\begin{pmatrix}
		0 & 0  \\
		0 & \Gamma_{B}(\omega, \vec{r}_{2})
		\end{pmatrix}$ respectively. $\Gamma_{A,B}$ is the linewidth function denoting the probe coupled to $A/B$ sublattices.  $\mathbf{G}^{M}$
is the retarded Green's functions of the sample including the probe effects. $\vec{r}_{1}$ and $\vec{r}_{0}$ are the positions of scanning and fixed probes respectively.
To simplify, assuming the effect of probes is weak, we neglect the influence of probes on sample Green's functions, then
\begin{equation}
	\begin{split}
	&T_{2B \leftarrow 1A}(\omega,\vec{r}_{1},\vec{r}_{0})\approx \Gamma_{A}(\omega)\Gamma_{B}(\omega) \vert G^{0,M}_{BA}(\omega,\vec{r}_{1}-\vec{r}_{0}) \vert^{2}\\
	&=\Gamma_{A}\Gamma_{B} \vert G^{0,M}_{BA}(\omega,\vec{r}) \vert^{2}\\
	\label{Eq6}
	\end{split}
\end{equation}
Here $\vec{r}=\vec{r}_{1}-\vec{r}_{0}$ is the relative distance vector between two probes. Using the Fourier transform, we can obtain the Fourier components $T_{2B \leftarrow 1A}(\omega, \vec{k})$. Specially, at the $\vec{k}=\vec{K}-\vec{K'}+\vec{q}=\Delta\vec{K}+\vec{q}$ is ($\vec{q}$ is a small number),
\begin{equation}
	\begin{split}
		&T_{2B\leftarrow1A}(\omega, \vec{k}=\Delta \vec{K}+\vec{q})\\
		&=\iint d^{2}re^{-i(\Delta \vec{K}+\vec{q}) \cdot \vec{r}}\Gamma_{A}\Gamma_{B} \vert G^{0,M}_{BA}(\omega,\vec{r}) \vert^{2}\\
		&=\iiiint \frac{d^{2}k d^{2}k'}{(2\pi)^{4}}\iint d^{2}re^{i(-\Delta \vec{K}-\vec{q}-\vec{k'}+\vec{k})\cdot \vec{r}}\Gamma_{A}\Gamma_{B}\\
		&\hspace{10mm} \times G^{0,M}_{BA}(\omega,\vec{k})
		G^{0,M*}_{BA}(\omega,\vec{k'})\\
		&\approx \iiiint \frac{d^{2}q_{1}d^{2}q_{2}}{(2\pi)^{4}}\iint d^{2}re^{i(-\vec{q}+\vec{q}_{1}-\vec{q}_{2})\cdot \vec{r}}\Gamma_{A}\Gamma_{B}\\
		&\hspace{10mm} \times G^{0,M}_{BA}(\omega,\vec{K}+\vec{q}_{1})
		G^{0,M*}_{BA}(\omega,\vec{K'}+\vec{q}_{2}).
	\end{split}
	\label{Eq7}
\end{equation}
The last approximation is because the main Fourier components of Green's functions focus on regions near the two valleys when the energy is relatively low. In the fast Fourier transform map of the STM experiment, $T_{2B \leftarrow 1A}(\omega,\vec{k}=\Delta \vec{K}+\vec{q})$ should correspond to peaks around the intervalley interference points $\vec{k}=\Delta \vec{K}$. Similar to previous references \cite{Dutreix,Zhang}, we do Fourier-filtering by only retaining the values of $T_{2B \leftarrow 1A}(\omega,\vec{k})$ at the points around $\Delta \vec{K}$ and discarding the other Fourier components. Then we do the inverse Fourier transform from them to obtain the Fourier transformed filtered transmission coefficient $T_{2B \leftarrow 1A} (\omega, \Delta \vec{K}, \vec{r})$,
\begin{equation}
	\begin{split}
    &T_{2B \leftarrow 1A} (\omega, \Delta \vec{K}, \vec{r})\\
	& = \iint \frac{d^{2}q}{(2\pi)^{2}}e^{i(\Delta \vec{K}+\vec{q}) \cdot \vec{r}}T_{2B \leftarrow 1A}(\omega, \vec{k}=\Delta \vec{K}+\vec{q})\\
    &\approx \Gamma_{A}\Gamma_{B}G^{0,M}_{K,BA}(\omega, \vec{r})G^{0,M*}_{K',BA}(\omega,\vec{r}),
    \end{split}
	\label{Eq8}
\end{equation}
where we use the relation $G^{0,M}_{K/K',BA}(\omega,\vec{r}) = \iint \frac{d^{2}q}{(2\pi)^{2}} e^{i (\vec{K}/\vec{K'}+\vec{q}) \cdot \vec{r}} G^{0,M}_{BA}(\omega, \vec{K}/\vec{K'}+\vec{q})$. With the same operation, the Fourier transformed filtered transmission coefficient $T_{2B \leftarrow 1A} (\omega, -\Delta \vec{K}, \vec{r})$ corresponding to $\vec{k}=-\Delta \vec{K}$ can be obtained by just switching $K$ and $K'$,
\begin{equation}
	\begin{split}
    T_{2B \leftarrow 1A} (\omega,-\Delta \vec{K}, \vec{r})
	\approx \Gamma_{A}\Gamma_{B}G^{0,M}_{K',BA}(\omega, \vec{r})G^{0,M*}_{K,BA}(\omega,\vec{r}).
    \end{split}
	\label{Eq9}
\end{equation}
Substituting the bare Green's functions in Eq.~(\ref{Eq4}) into Eq.~(\ref{Eq8}) and Eq.~(\ref{Eq9}), the final intervalley filtered transmission coefficients we aim to obtain is
\begin{equation}
	\begin{split}
    &T_{2B \leftarrow 1A}^{f}(\omega, \vec{r})=T_{2B \leftarrow 1A} (\omega, \Delta \vec{K}, \vec{r})+T_{2B \leftarrow 1A} (\omega, -\Delta \vec{K}, \vec{r})\\
    &=2\Gamma_{A}\Gamma_{B}Re\left[G^{0,M}_{K,BA}(\omega, \vec{r})G^{0,M*}_{K',BA}(\omega,\vec{r})\right]\\
	&\approx -2\Gamma_{A}\Gamma_{B}\left\vert \frac{\omega}{4\hbar^{2} v_{f}^{2}}H_{1}(\frac{\omega r}{\hbar v_{f}})\right\vert^{2} \cos[(\vec{K}-\vec{K}')\cdot \vec{r}+2\theta_{r}].
    \end{split}
	\label{Eq10}
\end{equation}
Analogically, if the fixed probe is connected to the sublattice $A$ and the scanning probe scans the sublattice $A$, the intervalley filtered transmission coefficient is approximated as
\begin{equation}
	\begin{split}
	T_{2A \leftarrow 1A}^{f}(\omega,\vec{r})\approx 2\Gamma_{A}^{2}\left\vert \frac{\omega}{4\hbar^{2} v_{f}^{2}}H_{0}(\frac{\omega r}{\hbar v_{f}})\right\vert^{2} \cos[(\vec{K}-\vec{K}')\cdot \vec{r}].
	\end{split}
	\label{Eq11}
\end{equation} 	
In the experiment, we apply a small voltage on the scanning probe 1 and ground the fixed probe 2. Then we measure the current flowing into the fixed probe 2. The differential conductance could be calculated by Landauer-B$\ddot{u}$ttiker formula $\frac{dI}{dV}=-\frac{2e^{2}}{h} T_{2 \leftarrow 1}$ at the zero temperature limit \cite{Datta}. Thus in the following we only pay attention on the transmission coefficients map.
Here the transmission coefficients or conductances exhibit a spatial oscillating behaviour via $\Delta \vec{K}=\vec{K}-\vec{K'}$ dependence. In Friedel oscillations, such a LDOS oscillation stems from the intervalley scatterings of electrons by vacancies or impurities \cite{Friedel,Dutreix}. In the present transport device, the spatial oscillating behaviour
comes from the intervalley interference on electron propagating paths [see Fig.~\ref{FIG1}(b)]. In addition,
the propagation between A sublattice to B sublattice involves a relative phase  $-\xi \theta_{q}$ in the momentum space and $-\xi \theta_{r}$ in the real space [see Eqs.~(\ref{Eq2}) and (\ref{Eq4})]. Such a relative phase originates from the fact that a gauge-invariant effective flux quantum always exist between A and B sublattice in the massless Dirac equation, whose sign is related to valley index $\xi$ \cite{Dutreix}.
It also reflects the distinct pseudospin-momentum locking texture for $K$ and $K'$ valley as shown in Fig.~\ref{FIG1}(c). Thus,
similar to the Friedel oscillations \cite{Dutreix,Zhang3}, an additional phase enters the transmission coefficient $T_{2B \leftarrow 1A}^{f}$ in Eq.~(\ref{Eq10}) as the twice of real space polar angle $\theta_{r}$.
The phase $\phi (r)=\Delta K \cdot r+2\theta_{r}$ acts as a potential field
whose gradient is the sum of a uniform field and a vortex \cite{Nye},
which is singular at $\vec{r}=0$.
Considering the transmission coefficient is a single-valued function,
it must return to the same value after circulating this field along a closed path. Thus, the path enclosing the field vortex should
contribute to a phase quantized to a topological number $2\pi N$ with an integer $N$. For $2 \theta_{r}$, the phase accumulation in an anticlockwise direction is $2 \pi N=4\pi$,
corresponding to 4 times the Berry phase $\gamma=\pi$ of the Dirac cone
and 2 additional wavefronts. Conversely, no additional phase accumulation and wavefronts appear for $T^f_{2A \leftarrow 1A}$, because no phase difference appears in the diagonal terms in Eq.~(\ref{Eq4}).

\begin{figure*}
	\includegraphics[width=1\textwidth]{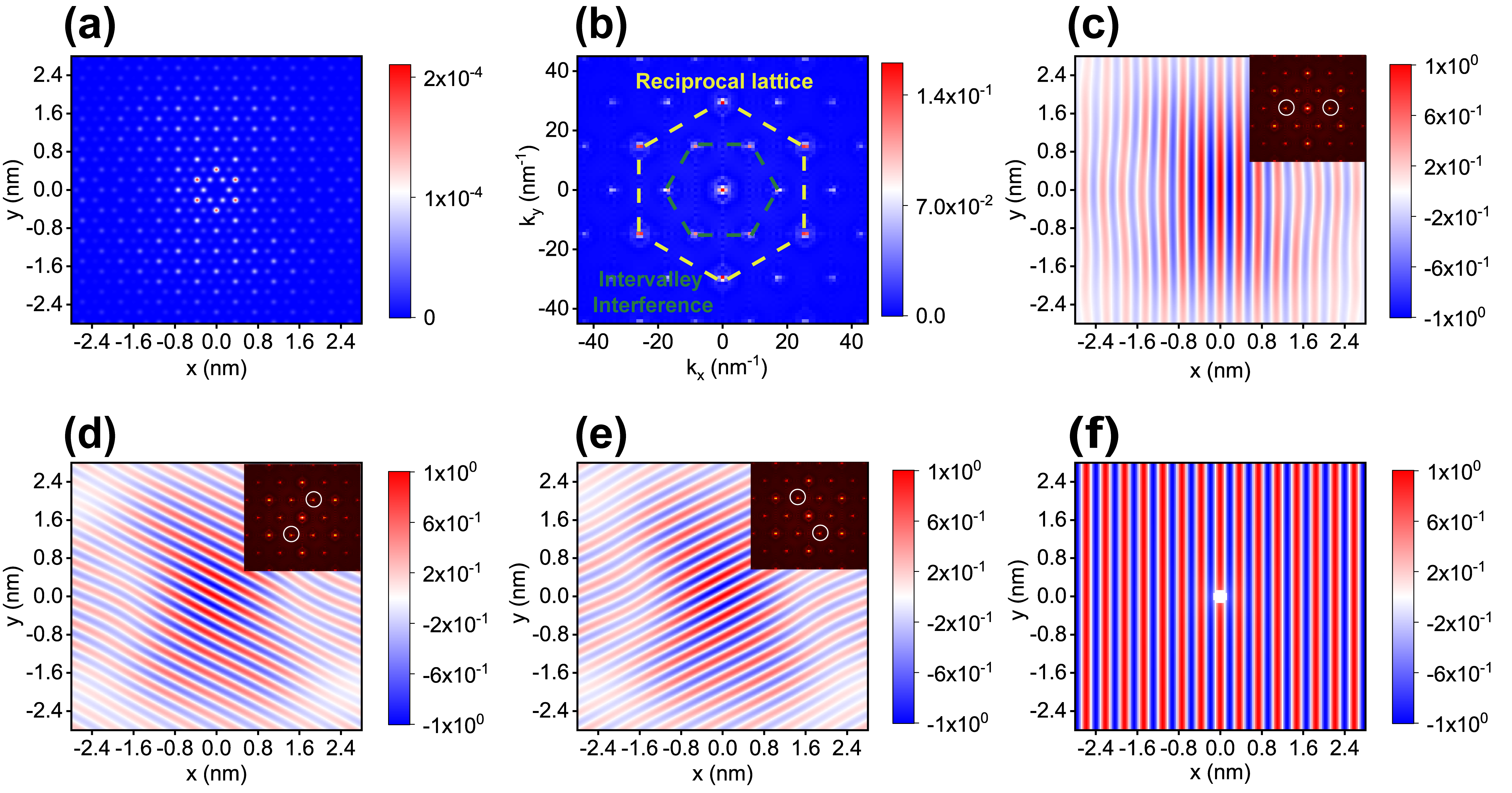}
	\centering 
	\caption{(a) The numerical calculated transmission coefficient map of $T_{2A \leftarrow 1A}(\omega)$. Note, when the fixed probe and scanning probe are located at the same A sublattice, the $T_{2A \leftarrow 1A}(\omega)$ is not estimated.  (b) Modulus of the FFT of the image in (a). The yellow and green dashed lines correspond to reciprocal lattices and the intervalley interference respectively. (c-e) The FFT-filtered images $T_{2A \leftarrow 1A}^{f}(\omega)$ of (b) schematically marked by white circles in the insets for three different directions.
	(f) The plot of the analytical transmission coefficient
	in Eq.~(\ref{Eq11}) which is multiplied by $r$. }
	\label{FIG4}
\end{figure*}

\subsection*{B. The results of numerical simulations}

To verify the above analytical analysis, we numerically calculate
the transmission coefficient and its fast Fourier transform (FFT)-filtered images by using
tight-binding model in this sub-section.
We construct a finite $N \times N$ graphene rectangular flake
[as shown in Fig.~\ref{FIG1}(a, b)]. Here $N$ is the number of lattice along one side of rectangular graphene flake (the number of sites along the transverse and longitudinal direction is the same). In the calculations, we choose $N=100$ corresponding to the width $W \approx 11nm$ and the length $L \approx 24nm$. The periodic boundary condition is applied (connecting the sites on the left and right boundaries, as well as the upper and lower boundaries respectively).
Its nearest-neighbor hopping tight-binding Hamiltonian is
\begin{equation}
	\mathbf{H}_{M}=-\sum_{<i,j>}t_{ij}c^{\dagger}_{i}c_{j}+h.c.
\end{equation}
See Fig.~\ref{FIG1}(a), assuming that the fixed (scanning) probes $1$ ($2$) connect with two sublattice $\imath$ ($\jmath$) respectively, their effects are quantized as the self-energy matrix
$\mathbf{\Sigma}^{r}_{1/2}$, where $(\mathbf{\Sigma}^{r}_{1})_{st}=
-i\frac{\Gamma_{1}}{2}\delta_{s\imath}\delta_{t\imath}$ or
$(\mathbf{\Sigma}^{r}_{2})_{st}
=-i\frac{\Gamma_{2}}{2}\delta_{s\jmath}\delta_{t\jmath}$.
The line-width matrix $\mathbf{\Gamma}_{1/2}=i(\mathbf{\Sigma}^{r}_{1/2}-\mathbf{\Sigma}^{r,\dagger}_{1/2})$. We also introduce self-energy terms $\Sigma_{D}^{r}=-i\Gamma_{D}/2$ on the lattices at the flake edges, to simulate an open environment on graphene brought about by electrode contacts on the boundaries in the experiment.
The retarded lattice Green's function including all effects is
\begin{equation}
	\mathbf{G}^{r,M}(\omega)=\frac{\mathbf{I}_{N\times N}}{(\omega+i0^{+})*\mathbf{I}_{N\times N}-\mathbf{H}_{M}-\mathbf{\Sigma}_{1}^{r}-\mathbf{\Sigma}_{2}^{r}-\mathbf{\Sigma}_{D}^{r}}
\end{equation}
Then, the transmission coefficient from probe $1$ to probe $2$ is numerically calculated as $T_{2 \leftarrow 1 }(\omega)=Tr[\mathbf{\Gamma}_{2}(\omega)
\mathbf{G}^{r,M}(\omega)\mathbf{\Gamma}_{1}(\omega)(\mathbf{G}^{r,M}(\omega))^{\dagger}]$\cite{Long,Sun}. In the calculation detail, $a=0.142nm$ is the length of carbon-carbon bond.
$t_{ij}=2.8eV$ (Fermi velocity $v_{f}=\frac{3t_{ij}a}{2\hbar} \approx 9*10^{5}m/s$), $\Gamma_{D}=0.8eV$ and $\Gamma_{1}=\Gamma_{2}=0.2eV$. The tip biased energy $\omega=1eV$. We here emphasize that our calculations are still in the low energy regime in view that $\omega<t_{ij}$ and the Fermi momentum $q_{F}=\omega/\hbar v_{f} \approx 1.7nm^{-1}$, which is much smaller than the intervalley distance $|\vec{K}-\vec{K'}| = \frac{4\pi}{3\sqrt{3}a}\approx 17nm^{-1}$.
After obtaining the transmission coefficients at each site,
we introduce a space broadening $\lambda_{d}$ by
$T_{2 \leftarrow 1}(\omega, \vec{r})=\sum_{\jmath}T_{2\jmath \leftarrow 1\imath}(\omega)e^{-\vert \vec{r}-\vec{r}_{\jmath} \vert^{2} / \lambda_{d}^{2}}$, where $\vec{r}$ denotes the space position of
the scanning tip $2$ center relative to the origin determined by the fixed tip $1$. Here $\lambda_{d}=0.05nm$.

Now we briefly describe our FFT filtering method which follows the same procedure in Refs. \cite{Dutreix,Zhang}. We first use the probe $2$ to scan a range of 6nm$\times$6nm
to obtain the transmission coefficient $T_{2 \leftarrow 1 }(\omega)$
(the conductance) in the real space [e.g. see Fig.~3(a)],
and then use the FFT to obtain the image in the reciprocal space [e.g. see Fig.~3(b)]. Some bright spots denoting the main Fourier components could be found. We pay attention to six intervalley interference points [i.e. bright dots circled by green dashed lines in Fig.~3(b)]. Next, we take the filtering in the reciprocal space,
that is, keep only the value in the white circles
and set to zero outer the white circles [e.g. see the insets
in Figs.~\ref{FIG3}(c-e)]. Note that two Fourier components with opposite momentum must be simultaneously included in the filtering process to ensure the FFT-filtered results are real, there will be three intervalley FFT-filtering directions [e.g. Figs.~\ref{FIG3}(c-e)]. Actually, these three directions reflect three intervalley interference processes since each $K$ point has three nearest-neighbour $K'$ point in the Brillouin. In view that the phase singularity in Eq.(10) is irrelevant to $\Delta \vec{K}$, they are expected to exhibit similar characteristics of wavefront dislocations. In the calculations, the diameter of white circles is set to around $7nm^{-1}$.
Finally, we take inverse FFT to obtain the oscillating transmission coefficient maps in the real space with the same size 6nm$\times$6nm [e.g. see Fig.~3(c-e)]. To avoid the inaccuracy on the boundaries, we focus on the range 5.6nm $\times$ 5.6nm. In addition, all the analytical or numerical calculated FFT-filtered transmission coefficients plotted in the figures (including the next sections) are modulated as $T^{f}/\max[T^{f}]$ where $\max[T^{f}]$ is the maximum value of $\vert T^{f}\vert $.

In Fig.~\ref{FIG3}(a), the fixed probe is connected to one $A$ sublattice and the scanning probe scans the surrounding $B$ sublattices.
In the numerical calculated $T_{2B \leftarrow 1A} (\omega, \vec{r})$ map, the $T_{2B \leftarrow 1A} (\omega, \vec{r})$ is bright near
the fixed probe at the origin
and gradually weakens as the distance becomes farther away.
This correlates with the decay relation implied by Hankel function of the first kind in Eq.~(\ref{Eq10}).
The modulus of the FFT of this transmission coefficient map
is shown in Fig.~\ref{FIG3}(b).
The bright spots in the FFT map indicates the main Fourier components \cite{Dutreix,Mallet}. The yellow dashed lines connects the bright spots at the reciprocal lattices which reflect the information of graphene lattices.
In addition, bright spots originated from the intervalley interference are connected by the green dashed lines.
They are surrounded in a hexagon with a side length $1/\sqrt{3}$ of that for reciprocal lattices. At the center, a bright ring denotes the intravalley scatterings, whose radius is roughly $2q_{F}$. Some discrepancies may be due to the scatterings and energy dissipations induced by the open environment at the graphene flakes boundaries.
We focus on the intervalley interference and do the filtering
around the white circles in the insets in Figs.~\ref{FIG3}(c-e)
and then take the inverse FFT.
The results show oscillating strip patterns with a wavelength of $\lambda_{\Delta K}=2\pi/\Delta K \approx 3.7 \mathring{A}$, which are just perpendicular to three FFT-filtered directions.
Clearly, two additional wavefronts appear near the origin,
where their positions are almost not affected by FFT-filtering directions as we expect.
It indicates a $4\pi$ phase accumulation related to a geometrical phase.
The number of the additional wavefronts is well consistent with
the analytical analysis in Eq.~(\ref{Eq10}),
plotted in Fig.~\ref{FIG3}(f) with $\Delta \vec{K}=(-\frac{4\pi}{3\sqrt{3}a},0)$.

\begin{figure*}
	\includegraphics[width=1\textwidth]{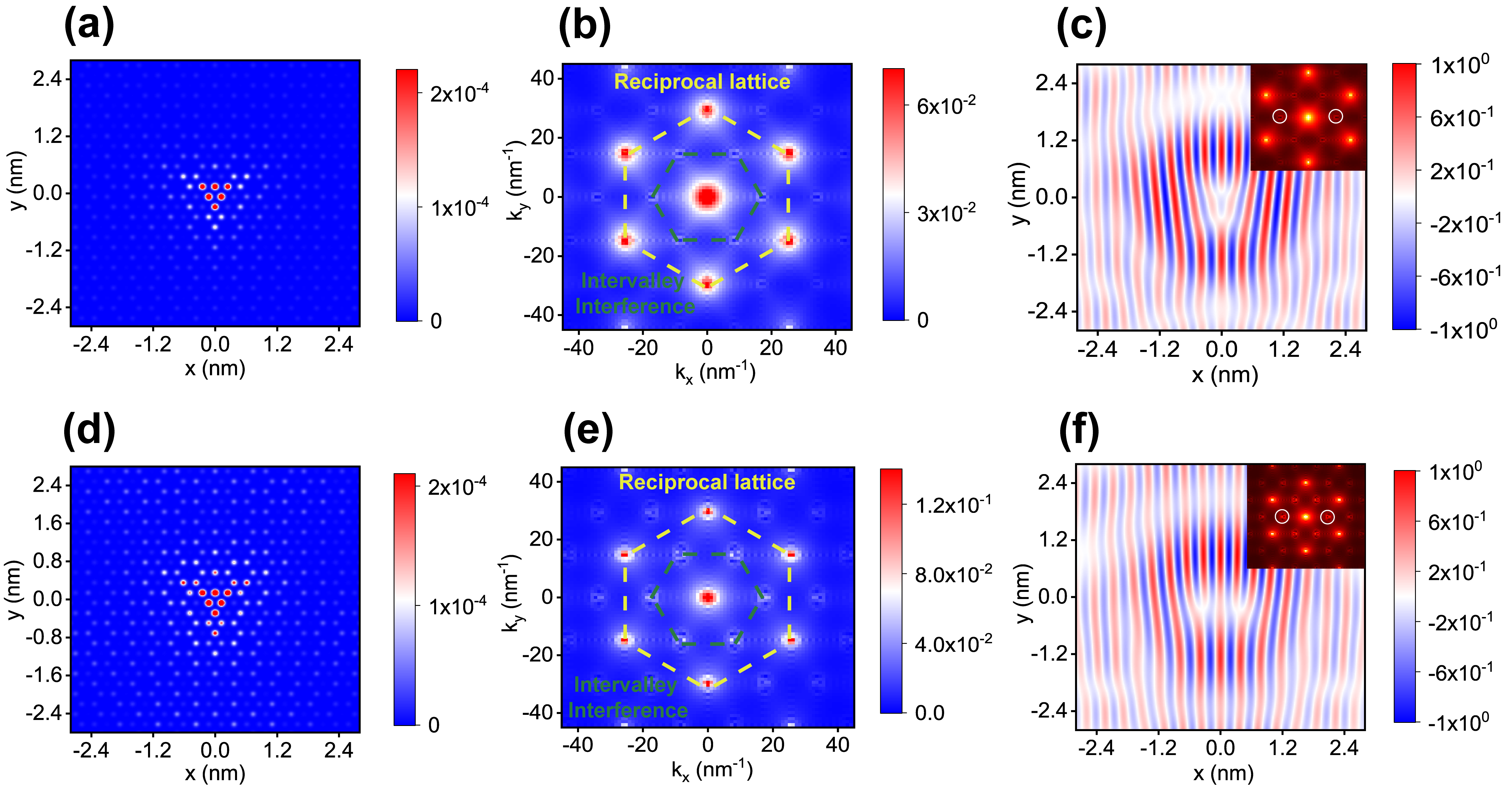}
	\centering 
	\caption{ The numerical calculated transmission coefficient map of $T_{2B \leftarrow 1A}(\omega)$ for a different tip biased energy and sample size for comparison.
Here (a) and (d) are the transmission coefficient map of $T_{2B \leftarrow 1A}$,
(b) and (e) are the modulus of the FFT of the image in (a) and (d),
and (c) and (f) are the FFT-filtered images. 
In panels (a-c), the tip biased energy $\omega=0.5eV$, and in (d-f), 
the sample size is $W \times L \approx 11nm \times 17nm$.
The other unmentioned parameters are same as Fig.~\ref{FIG3}.}
	\label{FIG5}
\end{figure*}

In Fig.~\ref{FIG4}(a), the fixed probe is selectively connected to one $A$ sublattice and the scanning probe scans the surrounding $A$ sublattices. Due to the change of scanned lattices, the transmission coefficient map shows peak positions slightly different from Fig.~\ref{FIG3}(a), but are still most bright near the center with $r=0$.
Because $A$ and $B$ lattices have the same periodic distributions, the FFT map of $T_{2A \leftarrow 1A} (\omega, \vec{r})$ in Fig.~\ref{FIG4}(b) still contributes main Fourier components at reciprocal lattices (connected by yellow dashed lines) and intervalley interference points (connected by green dashed lines). However, after we do inverse FFT at the intervalley points circled by white circles in Figs.~\ref{FIG4}(c-e), although the periodic oscillating strips appear as well, no wavefront dislocations exhibit for all three filtering directions. It means that encircling the intervalley interference points no longer threads an additional geometric phase for transmission coefficients between the same kind sublattices. This is consistent with the analytical analysis in Eq.~(\ref{Eq11}) plotted in Fig.~\ref{FIG4}(f).

\begin{figure*}
	\includegraphics[width=2\columnwidth]{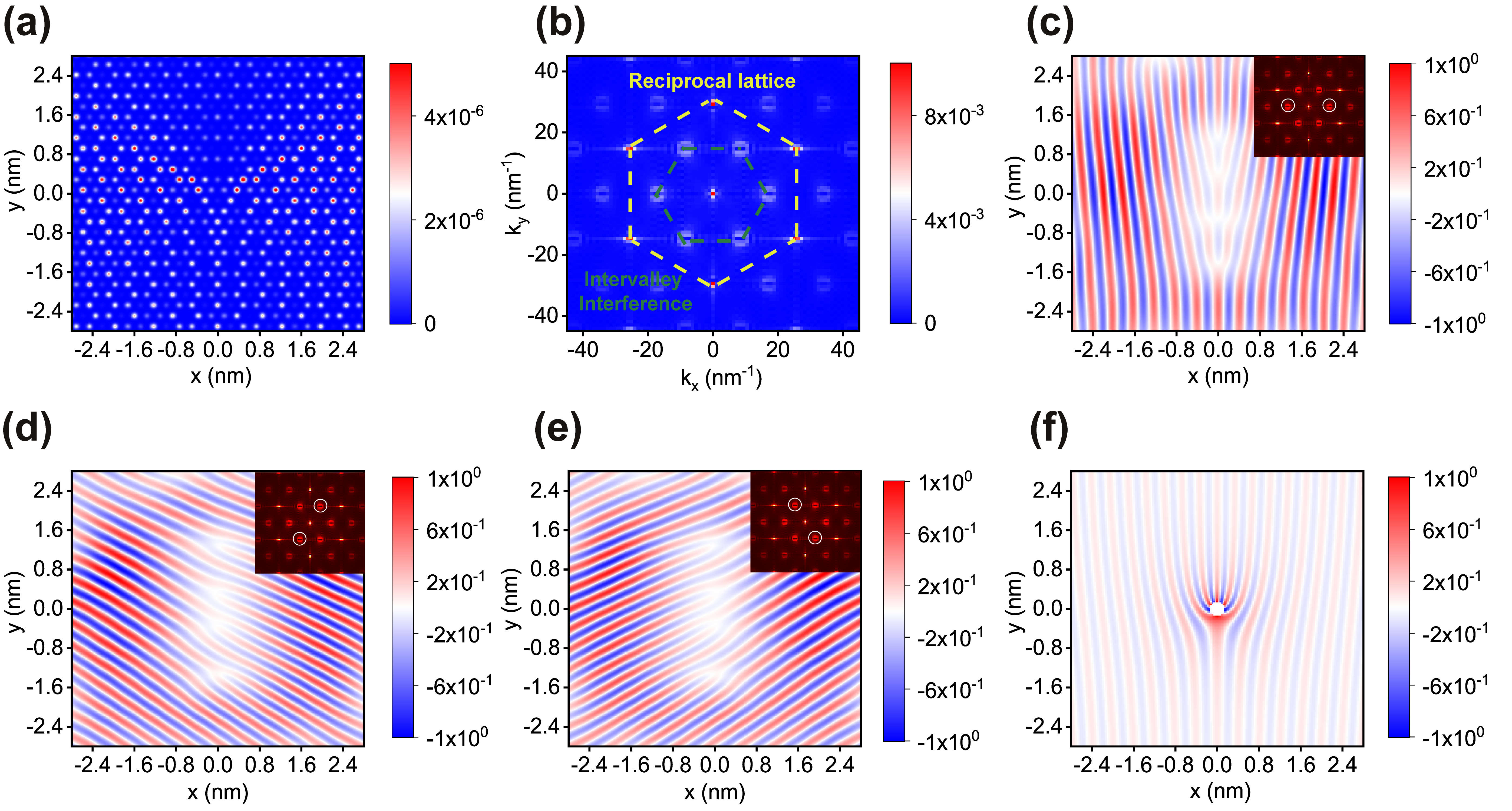}
	\centering 
	\caption{(a) The numerical calculated transmission coefficient map of $T_{2B \leftarrow 1A'}(\omega)$. (b) Modulus of the FFT of the image in (a). The yellow and green dashed lines correspond to reciprocal lattice and the intervalley interference respectively. (c-e) The FFT-filtered images $T_{2B \leftarrow 1A'}^{f}(\omega)$ of panel (b) schematically marked by white circles in the insets for three different directions.
	(f) The plot of the analytical transmission coefficient
	$T_{2B \leftarrow 1A'}^{f}(\omega)$ in Eq.~(\ref{Eq19}) which is multiplied by $r^{3}$.}
	\label{FIG6}
\end{figure*}

To investigate the effects of doping levels 
and sample sizes on our calculations, we present a numerical calculated 
transmission coefficient map of $T_{2B \leftarrow 1A}(\omega)$ 
for a different tip biased energy $\omega$ and sample size in Fig.~\ref{FIG5} 
to compared to Fig.~\ref{FIG3}. 
In Figs.~\ref{FIG5}(a-c), the tip biased energy $\omega=0.5eV$ 
and the sample size is the same as Fig.~\ref{FIG3}. 
In Figs.~\ref{FIG5}(d-f), we shorten the length of the graphene flake 
with $N=100$ sites along the longitudinal side and $M=70$ sites along the transverse side of the rectangle, then the size of the sample is $W \times L \approx 11nm \times 17nm$. 
We can find the variation of the tip biased energy and sample size could affect the scatterings and energy dissipations on the boundary, thereby influencing 
the intensity of intervalley scatterings and intravalley scatterings 
[Figs.~\ref{FIG5}(b,e)]. 
However, the number of wavefront dislocations in FFT-filtered results is still 2 
[see Figs.~\ref{FIG5}(c,f)], which is completely consistent with Fig.~\ref{FIG3}(c). 
This not only confirms the phase singularity is not affected by energy [Eq.~(\ref{Eq10})], 
but also confirms that our calculations have converged in size.

To conclude, the above results show that the phase singularities in the Green's functions will not only enter LDOS in Friedel oscillations \cite{Dutreix, Zhang3}, but also the transmission coefficients between different sublattices. The wavefront dislocations in the conductance map reflect the information of topological vortices. Enlightened by this, next we investigate the case for bilayer graphene.

\section{\label{sec3} The Bernal-stacked Bilayer graphene}
In this section, we generalize our theoretical setup to the Bernal-stacked bilayer graphene as shown in Fig.~\ref{FIG2}(a). Compared to monolayer graphene, two probes in the dual-probe STM experiment are connected to top and bottom layer of the bilayer graphene respectively. This separation makes the scanning of STM tips more convenient (see detailed discussions in Sec. IV). Moreover, we emphasize that the probe connected to the bottom graphene layer is not necessary to be a real STM tip. It could be replaced by a small electrode attached on the bottom sheet, serving as a drain in the transport measurement.

\subsection*{A. Analytical analysis of bilayer graphene}
For the Bernal-stacked bilayer graphene, we consider that the $A$ sublattices of the top graphene sheet are located on top of the $B'$ sublattices of the bottom graphene sheet. Then, the low-energy continuum Hamiltonian in the basis of $\{A,B,A',B'\}$ is \cite{Rozhkov,Nilsson}
\begin{equation}
	\begin{split}
		H_{\xi}^{D}(\vec{q}) &= \hbar v_{f}\begin{pmatrix}
			0 & \xi q_{x}-iq_{y} & 0 & t_{\bot }/\hbar v_{f} \\
		    \xi q_{x}+iq_{y} & 0 & 0 & 0\\
			0 & 0 & 0 & \xi q_{x}-iq_{y} \\
		    t_{\bot}/\hbar v_{f} & 0 & \xi q_{x}+iq_{y} & 0\\
			\end{pmatrix}
	\end{split}
\label{Eq14}
\end{equation}
where $t_{\bot}$ denotes the nearest-neighbour interlayer hopping. Here, the superscript D in the Hamiltonian as well as the following Green's functions denotes the bilayer (double layer) graphene. At the condition $t_{\bot} \gg 2\hbar v_{f} q$, the Hamiltonian can be further projected on the basis $\{B,A'\}$ \cite{Rozhkov}
\begin{equation}
	\begin{split}
		H_{\xi}^{D}(\vec{q}) \approx -\frac{\hbar^{2}v_{f}^{2}q^{2}}{t_{\bot}}\begin{pmatrix}
				0 & e^{2i \xi \theta_{q}} \\
				e^{-2i \xi \theta_{q}} & 0
	    \end{pmatrix}
	\end{split}
\label{Eq15}
\end{equation}
with the eigenvalues $E^{\pm}_{\xi}(\vec{q})=\pm \frac{\hbar^{2}v_{f}^{2}q^{2}}{t_{\bot}}$ and $\psi^{\pm}_{\xi}(\vec{q})=\frac{1}{\sqrt{2}}\binom{1}{\mp e^{-2i\xi \theta_{q}}}$. Similar to monolayer graphene, the eigenvectors also define pseudospin vectors $\langle \psi_{\xi}^{\pm}(q)\vert \vec{\sigma} \vert \psi_{\xi}^{\pm}(q)\rangle =\pm (-\cos 2\xi \theta_{q}, \sin 2\xi \theta_{q},0)$ lying on the $x-y$ plane with an azimuth angle $\pi - 2 \xi \theta_{q}$ in the Bloch space. Here $\sigma$ acts on the space expanded by the basis $\{B,A'\}$. This pseudospin-momentum locking texture is shown in Fig.~\ref{FIG2}(c). Particularly, after circulating a closed Fermi surface, the pseudospin rotates by $4\pi$, yielding a $W=-2\xi=\mp 2$ winding number and $\gamma=W\pi=\mp 2\pi$ Berry phase \cite{Park}.
The corresponding retarded Green's functions of
the bare bilayer graphene with the Hamiltonian in
Eq.~(\ref{Eq14}) can be calculated as
\begin{widetext}
\begin{equation}
	\begin{split}
		&\mathbf{G}_{\xi}^{0,D}(\omega,\vec{q})
		= \frac{1}{\Omega(\vec{q},\omega)}\times \\
		&\begin{pmatrix}
			\omega^{3}-\omega(\hbar v_{f}q)^{2} &\xi \hbar v_{f}q e^{-\xi \theta_{q}} [\omega^{2}-(\hbar v_{f} q)^{2}] & \omega t_{\bot} \xi \hbar v_{f}qe^{i\xi \theta_{q}} &\omega^{2}t_{\bot} \\ \xi
		    \hbar v_{f} q e^{\xi \theta_{q}} [\omega^{2}-(\hbar v_{f} q)^{2}] & \omega^{3}-\omega[(\hbar v_{f}q)^{2}+t_{\bot}^{2}] & t_{\bot} (\hbar v_{f}q)^{2}e^{i2\xi\theta_{q}}& \omega t_{\bot} \xi \hbar v_{f}qe^{i\xi \theta_{q}}\\
			\omega t_{\bot} \xi \hbar v_{f}qe^{-i\xi \theta_{q}}& t_{\bot} (\hbar v_{f}q)^{2}e^{-i2\xi\theta_{q}} & \omega^{3}-\omega [(\hbar v_{f} q)^{2} + t_{\bot}^{2}] & \xi \hbar v_{f}q e^{-i\xi \theta_{q}} [\omega^
			{2}-(\hbar v_{f} q)^{2}]\\ \omega^{2}t_{\bot} & \omega t_{\bot} \xi\hbar v_{f}qe^{-i\xi\theta_{q}} & \xi \hbar v_{f}q e^{i\xi \theta_{q}} [\omega^{2}-(\hbar v_{f} q)^{2}] & \omega^{3}-\omega(\hbar v_{f}q)^{2}
			\end{pmatrix} \\
        \label{Eq16}
	\end{split}
\end{equation}
\end{widetext}
where $\Omega(\omega,\vec{q})=(\hbar v_{f}q)^{4}+\omega^{2}[\omega^{2}-t_{\bot}^{2}-2(\hbar v_{f}q)^{2}]$. In the dual-probe STM measurements of bilayer graphene, we focus on four kinds of transmission coefficients $T_{2B \leftarrow 1A'}, T_{2A \leftarrow 1A'}, T_{2B \leftarrow 1B'}, T_{2A \leftarrow 1B'} $, which means the fixed probe $1$ is connected to one sublattice $A'$ or $B'$ of the bottom layer and the scanning probe $2$ selectively scans sublattices $A$ or $B$ of the top layer.
In analogy to Eqs.~(\ref{Eq8})-(\ref{Eq10}), their intervalley Fourier filtered terms are approximated as
\begin{equation}
	\begin{split}
    &T^{f}_{2B \leftarrow 1A'}(\omega)
	\approx 2\Gamma_{2B}\Gamma_{1A'}Re[G_{K,BA'}^{0,D}(\omega,\vec{r})G_{K',BA'}^{0,D*}(\omega,\vec{r})]\\
	&T^{f}_{2A \leftarrow 1A'}(\omega)
	\approx 2\Gamma_{2A}\Gamma_{1A'}Re[G_{K,AA'}^{0,D}(\omega,\vec{r})G_{K',AA'}^{0,D*}(\omega,\vec{r})]\\
	&T^{f}_{2B \leftarrow 1B'}(\omega)
	\approx 2\Gamma_{2B}\Gamma_{1B'}Re[G_{K,BB'}^{0,D}(\omega,\vec{r})G_{K',BB'}^{0,D*}(\omega,\vec{r})]\\
	&T^{f}_{2A \leftarrow 1B'}(\omega)
	\approx 2\Gamma_{2A}\Gamma_{1B'}Re[G_{K,AB'}^{0,D}(\omega,\vec{r})G_{K',AB'}^{0,D*}(\omega,\vec{r})].\\
	\end{split}
	\label{Eq17}
\end{equation}
By using Eq.~(\ref{Eq3}) and taking the approximation $\Omega (\vec{p},\omega) \approx (\hbar v_{f}q)^{4}-\omega^{2}t_{\bot}^{2}$ for $t_{\bot} \gg 2\hbar v_{f}q$ and $\omega$ \cite{Zhang},
the corresponding Green's functions of different valleys
$G_{K/K'}^{0,D}(\omega,\vec{r})$ which are Fourier transformed from the momentum space in Eq.~(\ref{Eq16})
to the real space are

\begin{equation}
	\begin{split}
	&G^{0,D}_{K,BA'}(\omega) \approx -\frac{it_{\bot}}{8(\hbar v_{f})^{2}}H_{2}(\frac{\sqrt{t_{\bot}\omega}r}{\hbar v_{f}})e^{i\vec{K} \cdot \vec{r}+i2\theta_{r}}\\
	&G^{0,D}_{K',BA'}(\omega) \approx -\frac{it_{\bot}}{8(\hbar v_{f})^{2}}H_{2}(\frac{\sqrt{t_{\bot}\omega}r}{\hbar v_{f}})e^{i\vec{K'} \cdot \vec{r}-i2\theta_{r}}\\
	&G^{0,D}_{K,AA'}(\omega)=G^{0,D}_{K,BB'}(\omega) \approx -\frac{\sqrt{t_{\bot}\omega}}{8(\hbar v_{f})^{2}}H_{1}(\frac{\sqrt{t_{\bot}\omega}r}{\hbar v_{f}})e^{i\vec{K} \cdot \vec{r}+i\theta_{r}}\\
	&G^{0,D}_{K',AA'}(\omega)=G^{0,D}_{K',BB'}(\omega) \approx \frac{\sqrt{t_{\bot}\omega}}{8(\hbar v_{f})^{2}}H_{1}(\frac{\sqrt{t_{\bot}\omega}r}{\hbar v_{f}})e^{i\vec{K'} \cdot \vec{r}-i\theta_{r}}\\
	&G^{0,D}_{K,AB'}(\omega)\approx \frac{i\omega}{8(\hbar v_{f})^{2}}H_{0}(\frac{\sqrt{t_{\bot}\omega}r}{\hbar v_{f}})e^{i\vec{K} \cdot \vec{r}}\\
	&G^{0,D}_{K',AB'}(\omega)\approx \frac{i\omega}{8(\hbar v_{f})^{2}}H_{0}(\frac{\sqrt{t_{\bot}\omega}r}{\hbar v_{f}})e^{i\vec{K'} \cdot \vec{r}}.\\
	\end{split}
	\label{Eq18}
\end{equation}
Substituting the Eq.~(\ref{Eq18}) into the Eq.~(\ref{Eq17}), the analytic forms of intervalley Fourier filtered transmission coefficients are
\begin{equation}
	\begin{split}
    &T^{f}_{2B \leftarrow 1A'}(\omega)\approx \\
	&\hspace{5mm}\frac{ \Gamma_{2B}\Gamma_{1A'}t_{\bot}^{2}}{32(\hbar v_{f})^{4}}\left\vert H_{2}(\frac{\sqrt{t_{\bot}\omega}r}{\hbar v_{f}})\right\vert^{2}\cos[\Delta \vec{K}\cdot \vec{r}+4\theta_{r}]\\
	&T^{f}_{2A \leftarrow 1A'}(\omega)\approx\\
	&\hspace{5mm}-\frac{ \Gamma_{2A}\Gamma_{1A'}t_{\bot} \omega}{32(\hbar v_{f})^{4}}\left\vert H_{1}(\frac{\sqrt{t_{\bot}\omega}r}{\hbar v_{f}})\right\vert^{2}\cos[\Delta \vec{K}\cdot \vec{r}+2\theta_{r}]\\
	&T^{f}_{2B \leftarrow 1B'}(\omega)\approx\\
	&\hspace{5mm}-\frac{ \Gamma_{2B}\Gamma_{1B'}t_{\bot} \omega}{32(\hbar v_{f})^{4}}\left\vert H_{1}(\frac{\sqrt{t_{\bot}\omega}r}{\hbar v_{f}})\right\vert^{2}\cos[\Delta \vec{K}\cdot \vec{r}+2\theta_{r}]\\
	&T^{f}_{2A \leftarrow 1B'}(\omega)\approx\\
	&\hspace{5mm}\frac{ \Gamma_{2A}\Gamma_{1B'}\omega^{2}}{32(\hbar v_{f})^{4}}\left\vert H_{0}(\frac{\sqrt{t_{\bot}\omega}r}{\hbar v_{f}})\right\vert^{2}\cos[\Delta \vec{K}\cdot \vec{r}].\\
	\end{split}
	\label{Eq19}
\end{equation}
As shown in Eq.~(\ref{Eq19}), these transmission coefficients also exhibit $\Delta \vec{K} =\vec{K}-\vec{K'}$  periodically oscillating behavior due to the valley interference on electron propagating paths,
see Fig.~\ref{FIG2}(b).  But different from the case of monolayer graphene, the phase singularities in Eq.~(\ref{Eq19}) suggest that $4,2,2,0$ additional wavefronts should appear near the origin in these $4$ kinds of transmission coefficients maps. The $4$ additional wavefronts are because the pseudospin vector of the bilayer graphene in Eq.~(\ref{Eq15}) could accumulate $4$ times Berry phase $\gamma = 2\pi$ along a path enclosing the origin. The $2$ and $0$ additional wavefronts can be understood as follows: the phase differences between $A$ ($A'$) and $B$ ($B'$) sublattices stay consistent with those in the monolayer graphene, since they are both distributed in the same top
(bottom) sheet. The interlayer hopping $t_{\bot}$ links the phase between $A$ sublattices and $B'$ sublattices. Based on this, the sublattices $B$ and $A'$, $A$ and $A'$ ($B$ and $B'$), $A$ and $B'$ could naturally possess a phase difference about $2\xi\theta_{q}$, $\xi\theta_{q}$ and $0$, as shown in Eq.~(\ref{Eq16}).
Due to the intervalley interference on the electron propagating paths,
it leads to the phase difference $4\theta_{r}$, $2\theta_{r}$ and $0$. $4,2,2,0$ additional wavefronts appear in the transmission coefficients
$T^{f}_{2B \leftarrow 1A'}$, $T^{f}_{2B \leftarrow 1B'}$,
$T^{f}_{2A \leftarrow 1A'}$, and $T^{f}_{2A \leftarrow 1B'}$,
respectively.
More generally, the above analysis can be extended to rhombohedral $l$-layer graphene, which is expected to show $N=2l$, $2l-2$, $...$, $0$ additional wavefronts due to $\gamma=l\xi\pi$ Berry phase \cite{Zhang}.

\subsection*{B. The results of numerical simulations}

The numerical simulations follow a similar procedure as the case for monolayer graphene in Sec. \ref{sec2}.B.
Here we consider a finite $N \times M$ bilayer graphene rectangular flake $N=200$ lattices (50 unit cells) along the longitudinal side and $M=50$ columns along the transverse direction of the flake. They correspond to the width $W \approx 11nm$ and length $L \approx 12nm$ respectively [Fig.~\ref{FIG2}(b)]. We also use the periodic boundary condition.
By using the tight-binding model, we numerically calculate
the transmission coefficient and its FFT-filtered images.
In the numerical calculation, we set $t_{\bot} = 0.4eV$ \cite{Zhang,Rozhkov}, and $\Gamma_{A}=\Gamma_{B}=\Gamma_{A'}=\Gamma_{B'}=0.2eV$. The tip biased energy $\omega=0.1eV$ corresponding to the Fermi momentum $q_{F}=\frac{\sqrt{\omega t_{\bot}}}{\hbar v_{f}} \approx 0.34nm^{-1}$, which is much smaller than the intervalley distance. This ensures our calculations in the low energy regime. All the other parameters are the same as Fig.~\ref{FIG3}(b).

In Fig.~\ref{FIG6}(a), the transmission coefficient map
of $T_{2B \leftarrow 1A'}(\omega)$ is shown, in which its strength still decays with the distance from the fixed probe. In the FFT map [Fig.~\ref{FIG6}(b)], we also find bright spots arranged as hexagons to denote main Fourier components for reciprocal lattices (connected by yellow dashed lines) and intervalley interference points (connected by green dashed lines) \cite{Zhang,Mallet}. Similar to the case of the monolayer graphene, a bright ring denoting intravalley scatterings also appear at the center.
In Figs.~\ref{FIG6}(c-e), we show the intervalley filtered inverse FFT map at the white circles in each inset.
Different from Fig.~\ref{FIG3}, four additional wavefronts emerge near the center for all three filtering directions, which indicates a 8$\pi$ phase accumulation surrounding the origin. As proved in Sec.~\ref{sec3}.A, this geometric phase comes from intervalley interference between distinct pseudospin phase difference $2\xi \theta_{r}$ in bilayer graphene. Note that the positions of wavefronts do almost not change with the filtering direction. In theory, four additional wavefronts should all emerge exactly at the origin, as shown in Fig.~\ref{FIG6}(f) for the analytic form of $T_{2B \leftarrow 1A'}^{f} (\omega)$ in Eq.~(\ref{Eq19}) with $\Delta \vec{K}=(-\frac{4\pi}{3\sqrt{3}a},0)$.
The deviation for the locations of additional wavefronts in Figs.~\ref{FIG6}(c-e) may be attributed to the boundary effect.

\begin{figure*}
	\includegraphics[width=2\columnwidth]{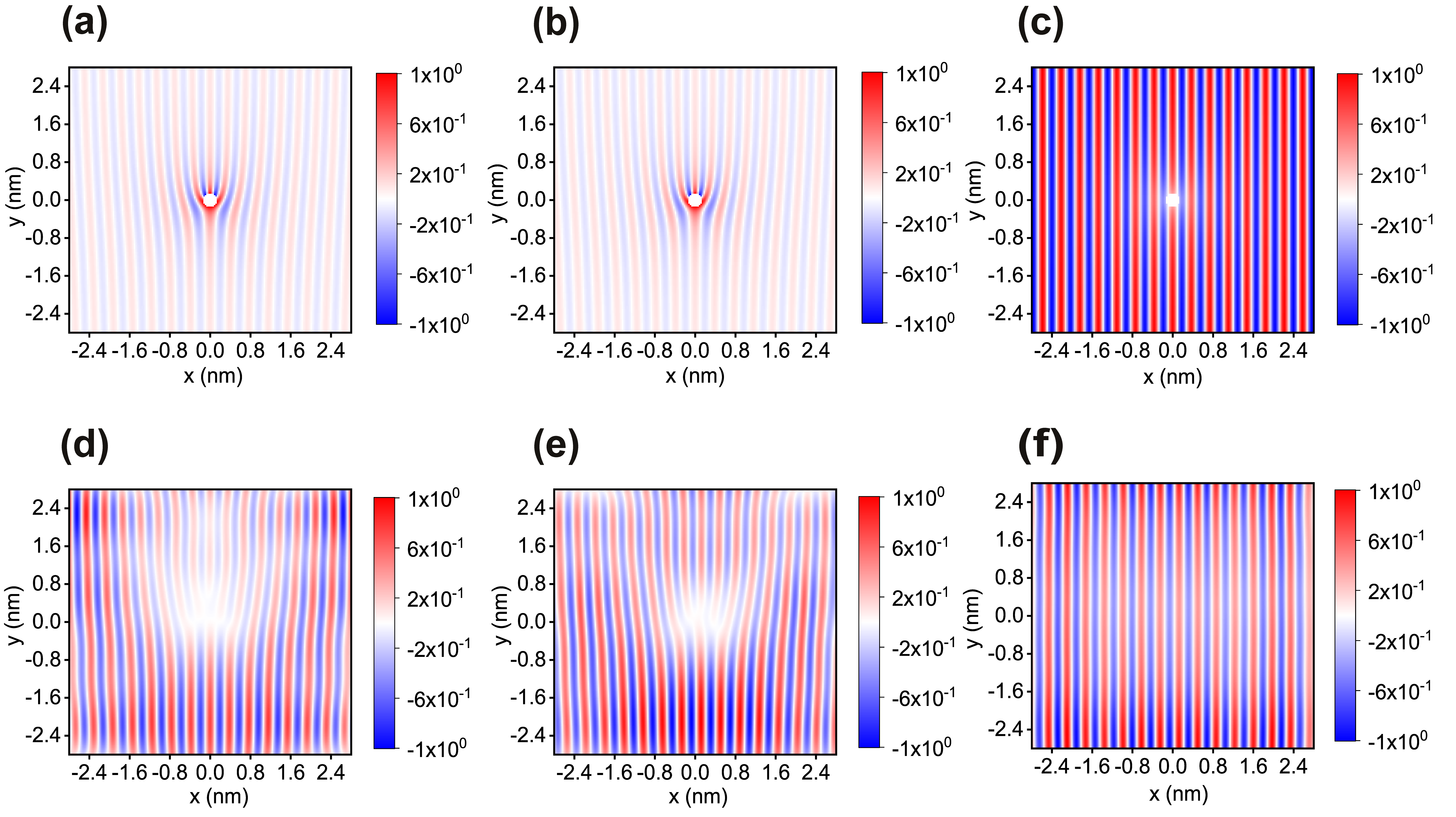}
	\centering 
	\caption{(a-c)
	The plot of the analytical transmission coefficients
	$T_{2A \leftarrow 1A'}^{f}(\omega)$, $T_{2B \leftarrow 1B'}^{f}(\omega)$ and $T_{2A \leftarrow 1B'}^{f}(\omega)$ in Eq.~(\ref{Eq19}) which is multiplied by $r$. (d-f) The numerical calculated FFT-filtered transmission coefficients $T^{f}_{2A \leftarrow 1A'}(\omega)$, $T^{f}_{2B \leftarrow 1B'}(\omega)$ and $T^{f}_{2A \leftarrow 1B'}(\omega)$ around the filtering points $\Delta \vec{K} = (\pm \frac{4\pi}{3\sqrt{3}a},0) $.}
	\label{FIG7}
\end{figure*}

Furthermore, the analytic forms of three transmission coefficients $T_{2A \leftarrow 1A'}$, $T_{2B \leftarrow 1B'}$ and $T_{2A \leftarrow 1B'}$ in Eq.~(\ref{Eq19}) with $\Delta \vec{K}=(-\frac{4\pi}{3\sqrt{3}a},0)$ are shown in Figs.~\ref{FIG7}(a-c). In comparison, the corresponding intervalley FFT-filtered numerical calculated transmission coefficients are shown in Figs.~\ref{FIG7}(d-f). Based on the symmetry, the results of $T_{2A \leftarrow 1A'}$ and $T_{2B \leftarrow 1B'}$ are basically same [see Eq.~(\ref{Eq19})]. Besides, two additional wavefronts arise, reflecting a $4\pi$ phase accumulation surrounding the origin. This proves the phase differences between $A$ ($B$) and $A'$ ($B'$) indeed contribute a geometric phase 2$\theta_{r}$. While for $T_{2A \leftarrow 1B'}$, no additional wavefronts exist, indicating the absence of the geometric phase. In total, all these characteristics confirm the validity of our analytical analysis.

\begin{figure}
	\includegraphics[width=1\columnwidth]{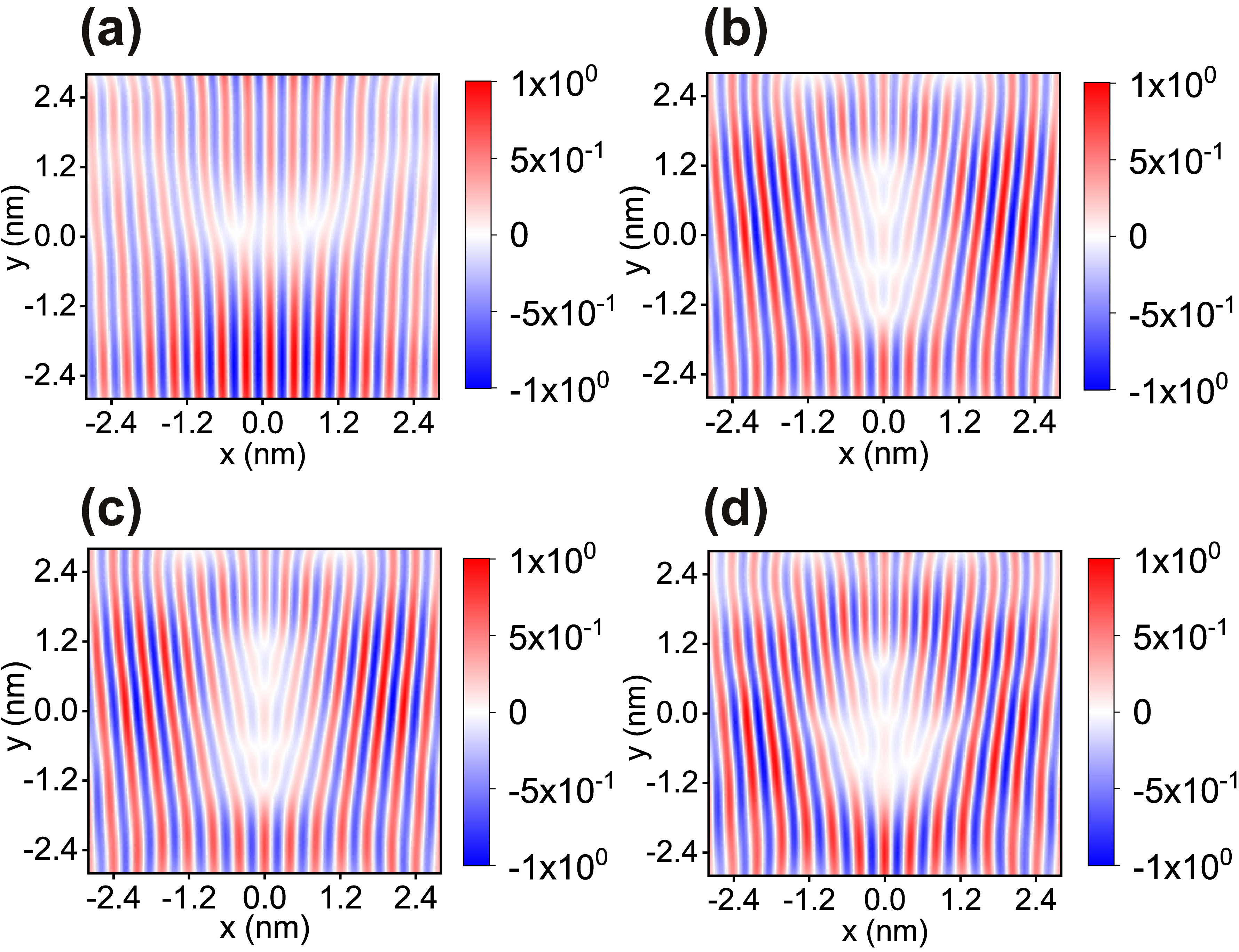}
   \centering 
   \caption{The numerical calculated FFT-filtered transmission coefficients (a) $T^{f}_{2(A+B)\leftarrow 1B'}(\omega)$, (b) $T^{f}_{2(A+B)\leftarrow 1A'}(\omega)$, (c) $T^{f}_{2(A+B)\leftarrow 1(A'+B')}(\omega)$ and (d) $\tilde{T}^{f}_{2(A+B)\leftarrow 1(A'+B')}(\omega)$ around the filtering points $\Delta \vec{K} = (\pm \frac{4\pi}{3\sqrt{3}a},0) $.}
   \label{FIG8}
\end{figure}

In a real scenario, because the STM tip has a finite size, it is difficult to couple to only one sublattice during the scanning process. But actually, even if the STM tip contact is not very ideal, our theoretical proposals in the bilayer graphene can still be applied. In Fig.~\ref{FIG8}, the fixed probe $1$ is connected with the $B'$ sublattice in Fig.~\ref{FIG8}(a) and $A'$ sublattice in Fig.~\ref{FIG8}(b), while the scanning probe $2$ scans all the sublattices surrounding the fixed probe. We denote these two transmission coefficients as $T^{f}_{2(A+B) \leftarrow 1B'}$ and $T^{f}_{2(A+B) \leftarrow 1A'}$. In the maps, $T^{f}_{2(A+B) \leftarrow 1B'}$ exhibits $2$ additional wavefront dislocations while $T^{f}_{2(A+B) \leftarrow 1A'}$ exhibits $4$ additional wavefront dislocations. Moreover, besides the scanning tip, the fixed tip (or the electrode) may also couple to multiple sublattices. For example, in Fig.~\ref{FIG8}(c), the fixed probe or the electrode is coupled to one $B'$ sublattice at the center and three surrounding $A'$ sublattices in the meanwhile (the coupling strength to $A'$ is set as half of the coupling strength to $B'$). This transmission coefficient is denoted as $T^{f}_{2(A+B)\leftarrow1(A'+B')}$ and still exhibits $4$ additional wavefront dislocations.  The above results are attributed to the fact that pseudospin components $B$ and $A'$ of the wave function play a leading role as a relatively low energy, see Eqs.~(\ref{Eq14}) and (\ref{Eq15}). This means $T^{f}_{2B \leftarrow 1A'}(\omega) \gg T^{f}_{2A \leftarrow 1A'}(\omega) \approx  T^{f}_{2B \leftarrow 1B'}(\omega) \gg T^{f}_{2A \leftarrow 1B'}(\omega)$ in our settings when $\omega \ll t_{\bot}$. So, although the probes may be coupled to multiple sublattices, $A'$ and $B$ sublattices are more important than $B'$ and $A$ sublattices. 

Furthermore, we also investigate the case that the fixed probe or the electrode 
is coupled to the next-nearest-neighbour A' sites.
In Fig.~\ref{FIG8}(d), the fixed probe 1 is coupled to one central A' site, 
its three nearest-neighbour B' sites and six next-nearest-neighbour A' sites 
on the bottom layer, while the scanning probe 2 scans all the lattices on the top layer (denoted as $\tilde{T}^{f}_{2(A+B) \leftarrow 1(A'+B')})$. 
The coupling strength of next-nearest-neighbour A' sites is a half of that of nearest-neighbour B' sites and a quarter of that of the central A' site. 
We can find that such a coupling situation does not destroy 4 additional wavefronts near the center [see Fig.~\ref{FIG8}(d)]. 
Thus, even if coupled to multiple lattices, our scheme in bilayer graphene could still be robust.
Of course, if the fixed probe or the electrode is very big to 
couple many carbon atoms, the expected results with 4 additional wavefronts 
may be buried. 
In terms of experimental realization, such a small electrode may be achieved by using a protrusion on the electrode to connect to a small area of the bottom sheet. 
It can be put right under the bilayer graphene, and covered by the insulating substrate. Under the substrate, the gating can adjust the doping level 
of the bilayer graphene \cite{Zhang6, Wong}. In total, we can expect that the dual-probe STM experiment is somewhat easier to achieve.

\subsection*{C. The biased bilayer graphene}
 In this subsection, to test the application range and robustness of our previous results, we further consider the biased bilayer graphene by applying a gate voltage in the out-of-plane direction. The electric field breaks the inversion symmetry and induces a continuously voltage-tunable band gap \cite{Castro2}. At the low energy range, the biased bilayer graphene with $2m$ biased energy can be described by Hamiltonian \cite{Rozhkov}:
 \begin{equation}
	\begin{split}
		H_{\xi}^{D}(\vec{q}) &= \hbar v_{f}\begin{pmatrix}
			m/\hbar v_{f} & \xi q_{x}-iq_{y} & 0 & t_{\bot }/\hbar v_{f} \\
		    \xi q_{x}+iq_{y} & m/\hbar v_{f} & 0 & 0\\
			0 & 0 & -m/\hbar v_{f} & \xi q_{x}-iq_{y} \\
		    t_{\bot}/\hbar v_{f} & 0 & \xi q_{x}+iq_{y} & -m/\hbar v_{f}\\
			\end{pmatrix}.
	\end{split}
\label{Eq20}
\end{equation}
The bare retarded Green's function matrix
$\mathbf{G}_{\xi}^{0,D}(\omega,\vec{q}) =
(\omega-H_{\xi}^{D}(\vec{q})+i0^+)^{-1}$
can be got as
\begin{widetext}
	\begin{equation}
		\begin{split}			
	&\mathbf{G}_{\xi}^{0,D}(\omega,\vec{q})=\frac{1}{\widetilde{\Omega}(\vec{q},\omega)}
			\begin{pmatrix}
				\omega_{-}[\omega_{+}^{2}-\tilde{q}^{2}] &\xi \tilde{q} e^{-\xi \theta_{q}} [\omega_{+}^{2}-\tilde{q}^{2}] & \omega_{-} t_{\bot} \xi \tilde{q}e^{i\xi \theta_{q}} &\omega_{-}\omega_{+}t_{\bot} \\ \xi
				\tilde{q} e^{\xi \theta_{q}} [\omega_{+}^{2}-\tilde{q}^{2}] & \omega_{+}[\omega_{+}\omega_{-}-t_{\bot}^{2}]-\omega_{-}\tilde{q}^{2} & t_{\bot} \tilde{q}^{2}e^{i2\xi\theta_{q}}& \omega_{+} t_{\bot} \xi \tilde{q}e^{i\xi \theta_{q}}\\
				\omega_{-} t_{\bot} \xi \tilde{q} e^{-i\xi \theta_{q}}& t_{\bot} \tilde{q}^{2}e^{-i2\xi\theta_{q}} & \omega_{-}[\omega_{-}\omega_{+}-t_{\bot}^{2}]-\omega_{+}\tilde{q}^{2} & \xi \tilde{q} e^{-i\xi \theta_{q}} [\omega_{-}^{2}-\tilde{q}^{2}]\\ \omega_{+}\omega_{-}t_{\bot} & \omega_{+} t_{\bot} \xi  \tilde{q}e^{-i\xi\theta_{q}} & \xi  \tilde{q} e^{i\xi \theta_{q}} [\omega_{-}^{2}-\tilde{q}^{2}] & \omega_{+}[\omega_{-}^{2}-\tilde{q}^{2}]
				\end{pmatrix} \\
			\label{Eq21}
		\end{split}
	\end{equation}
	\end{widetext}
where $\omega_{\pm}=\omega \pm m$, $ \tilde{q}=\hbar v_{f} q$
and $\widetilde{\Omega}(\vec{q},\omega) = \tilde{q}^{4}+\omega^{2}[\omega^{2}-2\tilde{q}^{2}-2m^{2}-t_{\bot}^{2}]+m^{2}[m^{2}+t_{\bot}^{2}-2\tilde{q}^{2}]$. It is easy to verify that once the biased energy $m=0$, Eq.~(\ref{Eq20}) and  Eq.~(\ref{Eq21}) directly restore to Eq.~(\ref{Eq14}) and Eq.~(\ref{Eq16}).
Comparing Eq.~(\ref{Eq21}) and Eq.~(\ref{Eq16}), the biased voltage $m$
(i.e. the mass term) does not break the phase correlations of the Green's function components, namely, $G^{0,D}_{\xi,BA'},G^{0,D}_{\xi,AA'},G^{0,D}_{\xi,BB'}$ and $G^{0,D}_{\xi,AB'}$ still have the phase $2\xi\theta_q$,
$\xi\theta_q$, $\xi\theta_q$ and $0$.
This can be further illustrated by projecting the $H_{\xi}^{D}(\vec{q})$
in Eq.~(\ref{Eq20}) into the subspace expanded by the basis $\{B,A'\}$ \cite{Rozhkov,Nilsson},
\begin{equation}
    \begin{split}
		H_{\xi}^{D}(\vec{q}) \approx \begin{pmatrix}
		m & -\frac{\hbar^{2}v_{f}^{2}q^{2}}{t_{\bot}}e^{2i \xi \theta_{q}} \\
		-\frac{\hbar^{2}v_{f}^{2}q^{2}}{t_{\bot}}e^{-2i \xi \theta_{q}} & -m
	    \end{pmatrix} .
    \end{split}
\end{equation}
Its eigenvalues are $E_{\pm}(\vec{q})=\pm \sqrt{F^{2}+m^{2}}$ and eigenvectors are $\psi^{\pm}_{\xi}(\vec{q})\propto (m \pm \sqrt{F^{2}+m^{2}}, -Fe^{-2i\xi \theta_{q}}$), where $F=\frac{\hbar^{2}v_{f}^{2}q^{2}}{t_{\bot}}$.
Although the pseudospin vectors defined by the eigenvectors $\langle \psi_{\xi}^{\pm}(q)\vert \vec{\sigma} \vert \psi_{\xi}^{\pm}(q)\rangle \propto  (-2F(m\pm \sqrt{F^{2}+m^{2}})\cos 2\xi \theta_{q}, 2F(m\pm \sqrt{F^{2}+m^{2}})\sin 2\xi \theta_{q},(m\pm\sqrt{F^{2}+m^{2}})^{2}-F^{2})$ are no longer pseudospin-momentum locking on the $x-y$ plane, they still have a $\pi-2\xi\theta_{q}$ azimuth angle.
Since the winding number is invariant here, the same phase accumulation around the origin and the number of additional wavefronts will appear
as the case for the biased energy $m=0$.

\begin{figure}
	\includegraphics[width=1\columnwidth]{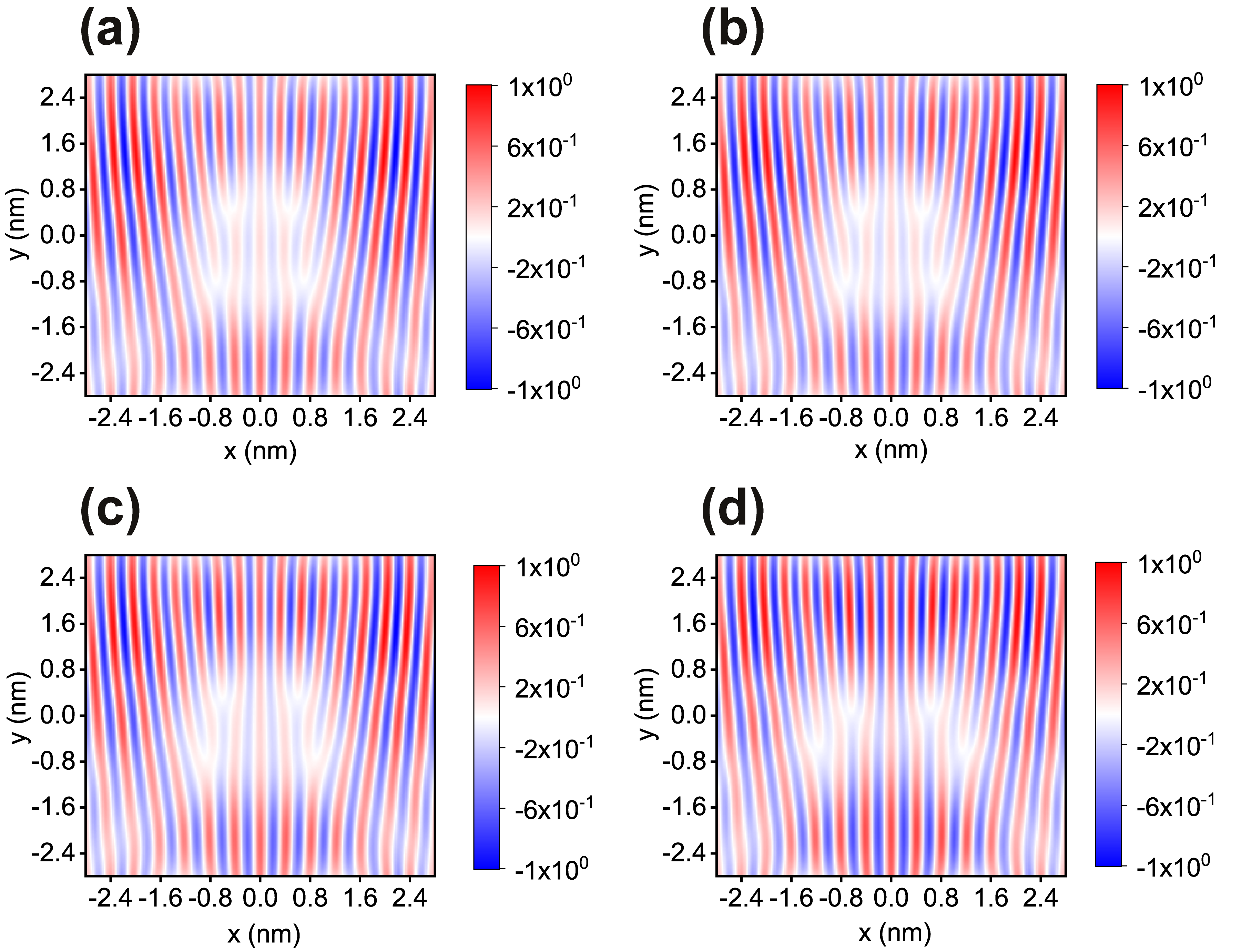}
   \centering 
   \caption{The numerical calculated FFT-filtered transmission coefficient $T_{2B \leftarrow 1A'}^{f}(\omega)$ around the filtering points $\Delta \vec{K} = (\pm \frac{4\pi}{3\sqrt{3a}},0) $ for the biased bilayer graphene. The biased energy (a) $m=0eV$, (b) $0.05eV$, (c) $0.1eV$, and (d) $0.15eV$.}
   \label{FIG9}
\end{figure}

In Fig.~\ref{FIG9}, we exhibit the numerical calculated FFT-filtered transmission coefficients map of $T_{2B \leftarrow 1A'}$. The energy $\omega=0.2eV$ and $t_{\bot}=0.4eV$. 
At first, the biased voltage is absent with $m=0$ [Fig.~\ref{FIG9}(a)], 
the result should be similar to Fig.~\ref{FIG6}(c) where $\omega=0.1eV$. 
We can find although the locations of wavefront dislocations change slightly, their number still keep four. This indicates our previous analysis does not depend on energy $\omega$, which is similar to the monolayer graphene.  Considering $\omega$ should exceed the energy gap $m$ to ensure the transport of the bulk states,  we increase the $m$ from $0eV$ to $0.15eV$ in Figs.~\ref{FIG9}(b-d). Apparently, the number of additional wavefronts remains to be four during the variation. Therefore, the wavefront dislocations are robust in biased bilayer graphene for a range of the biased voltage.  In fact, similar to the case of Friedel oscillations \cite{Zhang5,Yang}, our scheme could be applied to other gapped two-dimensional materials, like gapped graphene and transition metal dichalcogenides.

\section{\label{sec4} Summary and Discussion}

In this paper, we have raised a dual-probe scheme to demonstrate
the wavefront dislocations in monolayer and Bernal-stacked bilayer
graphene by the transport measurement.
The analytical analysis shows that due to the distinct pseudospin textures
at $K$ and $K'$ valleys, the intervalley interference on the electron
propagating paths could additionally introduce a phase singularity
into transmission coefficients between different sublattices.
This phase singularity acts as a topological defect
and contributes additional wavefronts related to topological indices on the conductance map. A comprehensive numerical tight-binding calculations combined with FFT analysis further support our conclusions. Especially for the bilayer graphene, the wavefront dislocations is found to still exist even if the tips are connected to multiple sublattices. Our scheme is also applied for biased bilayer graphene and some other gapped two-dimensional materials. Our work opens up a new transport routine on experiment to explore valley-related topological properties of materials.

Now we will discuss the details about experimental implementation.
In the current technical conditions, STM already allows imaging of the topography of surfaces, mapping the distribution of LDOS and manipulating individual atoms and molecules all at atomic resolutions \cite{Hansma,Ko}. Therefore, the coupling of one atom site for the STM tip is in principle possible to be achieved in practice.
The integration of several atomically precisely controlled probes in a multiprobe STM system has also continuously developed in recent years \cite{Nakayama, Ko}. To satisfy our scheme, the key point is the distance between the probes during the two-probe transport measurement.
As reported by Kolmer at al, in two-probe STM/STS (Scanning tunneling spectroscopy) experiments on about 70nm long dangling bond dimer wire supported on a hydrogenated Ge(001) surface \cite{Kolmer1} and the anisotropic Ge(001) surface \cite{Kolmer}, the probe to probe separation distance could be reduced to tens of nanometers.
The main limitation on the distance between STM probes is tip diameters. So, to achieve a smaller tip-tip distance, finer tips are required, which may cause some difficulties on implementation. However, such demanding condition are naturally relaxed in experiments on the bilayer graphene, since the two probes are spatially separated by the top and bottom graphene layers. Particularly, because the probes contact need not be limited on one site in bilayer graphene, the fixed probe connected to the bottom sheet as a source can also be replaced by a small electrode.

Photonic lattices \cite{Rechtsman,Polini,Wu} can also serve to implement our theoretical scheme. The photonic graphene is composed of a periodic array of evanescently-coupled waveguides in a honeycomb structure, which is often used to explore some fundamental wave-transport phenomena \cite{Schwartz,Rechtsman2}.  Because the distance between each waveguide commonly reaches several $\mu m$, the light intensity at each lattice is easier to be obtained. In fact, there is a study to demonstrate the pseudo-mediated vortex generation in artificial photonic graphene by three or two beam excitation \cite{Song}. For our proposal,
we can inject the light on one single waveguide of the input facet of the array, and measure the intensity profile at the output facet for desired sublattices. This process is equal to the finite-time evolution of an electron wavepacket and somehow similar to our proposed transport measurement.

\section*{Acknowledgments.}
This work was financially supported by NSF-China (Grant No. 11921005), the Innovation Program for Quantum Science and Technology (2021ZD0302403), and the Strategic Priority Research Program of Chinese Academy of Sciences (Grant No. XDB28000000).
\\
\\

\bibliography{ref}
\end{document}